\documentclass[linenumbers]{emulateapj}
\usepackage{aas_macros}
\usepackage{epstopdf}
\usepackage{epsfig}
\usepackage{natbib}
\usepackage{float}
\usepackage{gensymb}
\usepackage{amsmath}
\usepackage{lipsum}% http://ctan.org/pkg/lipsum
\usepackage{times}
\usepackage{comment}
\usepackage{color}
\usepackage{multirow}
\usepackage{soul}
\usepackage{CJKutf8}            % Chinese name
\usepackage{xspace}
\usepackage{lineno}

\definecolor{purple}{RGB}{160,32,240}

\definecolor{purple2}{RGB}{120,72,240}

\newcommand{\hi}{\ion{H}{1}\xspace}
\newcommand{\ovi}{\ion{O}{6}\xspace}
\newcommand{\ncode}[1]{{\sc #1}}
\newcommand{\cloudy}{\ncode{Cloudy}\xspace}

\begin{document}

%\title{Emission of the Ionized Gas in Group of Galaxies}
\title{An Empirical Determination of the Dependence of the Circumgalactic Mass Cooling  Rate and Feedback Mass Loading Factor on Galactic Stellar Mass}

\author{Huanian Zhang \begin{CJK*}{UTF8}{gkai}(张华年)\altaffilmark{1}, Dennis Zaritsky\altaffilmark{1},  Karen Pardos Olsen \altaffilmark{1},  Peter Behroozi \altaffilmark{1},  Jessica Werk \altaffilmark{2},  Robert Kennicutt \altaffilmark{1,3},  Lizhi Xie (谢利智) \altaffilmark{4}, Xiaohu Yang (杨小虎) \altaffilmark{5,6},  Taotao Fang (方陶陶) \altaffilmark{7} ,  Gabriella De Lucia \altaffilmark{8}, Michaela Hirschmann \altaffilmark{9}, Fabio Fontanot \altaffilmark{8,10}\end{CJK*}}
\altaffiltext{1}{Steward Observatory, University of Arizona, Tucson, AZ 85719, USA; fantasyzhn@email.arizona.edu}
\altaffiltext{2}{Department of Astronomy, University of Washington, Seattle, WA 98195,  USA}
\altaffiltext{3}{George P. and Cynthia W. Mitchell Institute for Fundamental Physics \& Astronomy, Texas A\&M University,  College Station, TX  77843-4242, USA}
 \altaffiltext{4}{Tianjin Normal University, Binshuixidao 393, 300387, Tianjin, China}
\altaffiltext{5}{Department of Astronomy, School of Physics and Astronomy, and Shanghai Key Laboratory for Particle Physics and Cosmology, Shanghai JiaoTong University, Shanghai, 200240, China}
\altaffiltext{6}{Tsung-Dao Lee Institute, and Key Laboratory for
    Particle Physics, Astrophysics and Cosmology, Ministry of Education, Shanghai Jiao Tong University 
  Shanghai, 200240, China}
  \altaffiltext{7}{Department of Astronomy, Xiamen University, Xiamen, Fujian, China }
 \altaffiltext{8}{ INAF - Astronomical Observatory of Trieste, via G.B. Tiepolo 11, I-34143 Trieste, Italy}
 \altaffiltext{9}{DARK, Niels Bohr Institute, University of Copenhagen, Niels Bohr Building, Jagtvej 128, 2200 Copenhagen, Denmark}
\altaffiltext{10}{IFPU - Institute for Fundamental Physics of the Universe, via Beirut 2, 34151, Trieste, Italy}

\begin{abstract}
Using our measurements of the H$\alpha$ emission line flux originating in the cool (T $\sim10^4$ K) gas that populates the halos of galaxies, we build a joint model to describe mass of the cool circumgalactic medium (CGM) as a function of galactic stellar mass ($10^{9.5} < ({\rm M_*/M}_\odot) < 10^{11}$) and environment. Because the H$\alpha$ emission correlates with the main cooling channel for this gas, we are able to estimate the rate at which the CGM cools and becomes fuel for star formation in the central galaxy. We describe this calculation, which uses our observations, previous measurements of some critical CGM properties, and modeling of the cooling mechanism using the \cloudy modeling suite. We find that the mass cooling rate  is larger than the star formation rates of the central galaxies by a factor of $\sim 4 - 90$, empirically confirming that there is sufficient fuel to resolve the gas consumption problem and that feedback is needed to avoid collecting too much cold gas in galaxies. 
We find excellent agreement between our estimates of both the mass cooling rates and mass loading factors and the predictions of independent theoretical studies. The convergence in results that we find from several completely different treatments of the problem, particularly at the lower end of the galactic mass range,
is a strong indication that we have a relatively robust understanding of the quantitative effects of feedback  across this mass range. 
\end{abstract}

\keywords{galaxies: kinematics and dynamics, structure, halos, intergalactic medium}

\section{Introduction}

The interstellar medium (ISM) alone can not sustain the star formation activity of a galaxy for its entire lifespan, thus necessitating the inflow of cool gas from the circumgalactic medium (CGM) to provide additional fuel \citep{spitzer}. Indeed, the COS-GASS survey found a correlation between interstellar and circumgalactic \hi, implying a physical connection between the \hi disk and the CGM such that the \hi disk is nourished by accretion of gas from the CGM \citep{borthakur2015}. Through observations and simulations, we have now arrived at the conclusion that the baryon cycling between the CGM and central galaxy is the dominant factor driving the observable properties of galaxies \citep[see][for a review]{CGM2017}. As such, measuring the net gas mass inflow rate is key to our understanding of galaxy evolution.

Hydrodynamic simulations have shed light on the delicate process of  precipitating cold gas out of the hot medium in the CGM. Small-box simulations of a region of a galactic wind flow or the CGM can reach the resolution needed to track cooling/precipitation and have been used to study the possible origin of phenomena such as cold molecular clumps found in quasar outflows \citep[e.g.][]{ferrara2016, gronke2018, liang2020}. Larger cosmological simulations provide the information required to understand more global phenomena such as the observed \ovi content of galactic haloes \citep[e.g.][]{suresh2017,oppenheimer2018, Peeples2019,vandeVoort2019,Hummels2019,Nelson2020}. Among others, one of the ultimate goals of these hydrodynamic simulations, combined with more phenomenological treatments, is to determine the inflow rate of cold gas into the central portion of galaxies and the fraction of that gas that must be removed or reheated, so that together they fuel and throttle the process of star formation on galactic scales \citep{Muratov2015,Hayward2017,xie2017,xie2020}.

 Among observational studies, most attempt to trace the CGM and establish its basic characteristics (e.g., total mass, radial density profile, metallicity, kinematics).
There are relatively few empirical constraints on the {\it cooling rate} itself which is more closely related with the star formation process in galaxies \citep[for one exception to this statement see][which models the edges of cold clouds to estimate the precipitation rate] {Voit2019}.  The reason for the absence of constraints on the cooling rate is that 
studies of the CGM have come primarily from the study of absorption lines in the spectra of bright background objects \citep[e.g.][]{steidel2010,Chen2010,menard2011,bordoloi2011,zhu2013a,zhu2013b,Werk2013,Johnson2013,Johnson2014,Werk2014,Johnson2015,werk16,croft2016,croft2018,prochaska2017,Cai2017,Johnson2017,Chen2010,Chen2017a,Chen2017b,lan2018,joshi2018,Chen2019,Zahedy2019}. However, a growing set of studies is focusing on emission lines. For example, there are now measurements of UV or soft X-ray emission lines in either low-redshift starburst galaxies \citep{Hayes2016} or high-redshift galaxies \citep{Vikhlinin2006, Cantalupo2014, Prescott2015, Cai2019}. 

In emission lines, we are directly viewing the escaping energy from the CGM.
As such, the measurement of emission lines originating from the CGM offers the potential opportunity to explore questions regarding the cooling rates. 
 Of course, the problem is more complex than any simple scenario. In the local Universe,  galactic coronae can be studied via their X-ray emission, but X-ray observations of the Milky Way and nearby galaxies have shown that charge exchange X-ray emission, generated at the interface between the cool and hot gas in the halo, may also make a non-negligible contribution to the cooling \citep[e.g.][]{henley2015,jiang2019}. As such, the reader should bear in mind that our discussion of the mass cooling process in the following is likely to be oversimplified.

Optical emission lines provide an opportunity to explore the cooling rate of the cool CGM, but are extremely challenging to measure. In the local universe, H$\alpha$ is detected in {\sl individual} galaxies only when the systems are extreme and difficult to model \citep[such as in the starburst/merger NGC 6240;][]{Yoshida2016}. 
However, \cite{zhang2016} presented the first detection of the emission line
flux of H$\alpha$ and [N{ \small II}] $\lambda$6583, from low redshift, normal galaxies extending out to $\sim$ 100 kpc projected radius by stacking a sample of over 7 million lines of sight from the Sloan Digital Sky Survey \citep[SDSS DR12;][] {SDSS12}. 
Building on that first result, we have applied the technique to characterize the nature of the emission of the CGM  within 50 kpc in low redshift galaxies \citep[][hereafter, Papers I, II, III, IV, V, VI]{zhang2016,zhang2018a,Zhang2018b, Zhang2019, Zhang2020a, Zhang2020b}. 
We restrict ourselves to the inner 50 kpc because the measured emission beyond that is strongly affected by emission from nearby galaxies (Paper II). In Paper IV and V, we studied and modeled the environmental dependence and the stellar mass dependence,  respectively.

Theoretical
studies in the literature \citep[e.g.,][]{Mathews1978,Schure2009,Lykins2013, Vasiliev2013} have examined the cooling of  the plasma within the temperature range of $10^4 - 10^8$ K,  which can be assumed to be in collisional ionization equilibrium, using the hydrogen and metal line emission fluxes of the elements H, He, C, N, O, Ne, Mg, Si and Fe for temperatures of $10^4 - 10^8$ K, among which the hydrogen emission flux can be dominant for the temperatures of $10^4 - 10^5$ K. These existing, extensive studies of cooling rates and the computational infrastructure developed to study the ISM and CGM, such as \cloudy \citep{Cloudy98,Cloudy13, Cloudy17}, are key to our study.

We aim to convert the observed H$\alpha$ emission flux radiating from the CGM of galaxies into an estimate of the mass cooling rate in units of M$_\odot$ yr$^{-1}$. We will then use this information to infer the required rate at which infalling gas must be reheated or ejected back into the CGM, under the hypothesis of steady state equilibrium, by comparing the mass cooling rate to the star formation rate. This will be compared to theoretical estimates of these quantities derived from requiring a match between other observable properties of the galaxies (stellar mass, star formation rates) in their models. Multiple approaches in estimating these quantities are critical given the likely presence of systematic errors in any individual approach. Before doing all of this, we revisit  previous empirical results and models of the CGM dependence on stellar mass and environment using both an expanded data set and a slightly more complicated model to ensure that we have the latest estimates of CGM properties with which to estimate the mass cooling rate and feedback mass loading factor. 

The paper is organized as follows. In \S\ref{sec:dataAna} we present the data analysis and results, with a focus on how the emission flux depends on galactic stellar mass and local environment. In \S\ref{sec:model} we discuss our joint modeling of the CGM cool gas mass fraction on both galactic stellar mass and local environment. We describe our calculation of the cooling rate in \S\ref{sec:cooling rate}, including the application of \cloudy in \S\ref{sec:cloudy}. We present our estimates of the mass cooling rates and mass loading factors  in \S\ref{sec:mcr}. Lastly, we summarize in \S\ref{sec:sum}.
Throughout this paper, we adopt a $\Lambda$CDM cosmology with parameters
$\Omega_m$ = 0.3, $\Omega_\Lambda =$ 0.7, $\Omega_k$ = 0 and the dimensionless Hubble constant $h = $ 0.7 \citep[cf.][]{riess,Planck2018}.

\section{Data Analysis}
\label{sec:dataAna}

We follow the approach developed in Papers I through VI \citep[][ respectively]{zhang2016,zhang2018a,Zhang2018b,Zhang2019,Zhang2020a,Zhang2020b}, by selecting galaxies that meet our criteria in redshift ($0.02 < z < 0.2$),  half light radius ($1.5 < R_{50}/{\rm kpc} < 10$),  and  luminosity ($10^{9.5} < L /L_\odot < 10^{11}$) as primary galaxies.  For these galaxies, we extract measurements of the S\'ersic index ($n$) and $r$-band absolute magnitude ($M_r$) from \citet{simard}, stellar mass (M$_*$) from  \cite{Kauffmann2003a,Kauffmann2003b} and \cite{Gallazzi}, and star formation rates (SFR) from the MPA-JHU catalog \citep{Brinchmann}.  The SFR estimates are aperture corrected to account for the light outside the SDSS fiber aperture, which only collects $\sim$ 1/3 of the total light for a typical galaxy at the median redshift of the
survey \citep[for details see ][]{Brinchmann}. Our requirements for these ancillary measurements limit the primary sample to galaxies from the 7th major data release of SDSS (DR7), but we utilize spectra from DR16 to probe the CGM of these galaxies.

We  collect the spectra from the SDSS DR16 \citep{SDSSDR16} for all lines-of-sight projected within a specified range of scaled projected radii from any of our primary galaxies. As described in Paper V, 
we use scale projected radii to account for the range in primary galaxy sizes. We define the scaled projected radius, $r_s$, as the ratio between the physical projected separation and the virial radius of the primary galaxy ($r_s  \equiv r_p/r_{\rm vir}$). To estimate the virial radius of the primary galaxy, we use the scaling relation between the luminosity and the virial radius obtained by fitting a high order polynomial to the results drawn from the UniverseMachine \citep{Behroozi2018}. As discussed in more detail in Paper III, 
we set a physical lower limit on the projected radius (10 kpc) we explore to mitigate possible contamination of the line of sight spectra by the central  galaxy. We confirm, but discuss no further below, that increasing this lower limit on the projected separation to 20 kpc produces statistically consistent results  except for those involving the innermost radial bin of the smallest primary galaxies. Those measurements are affected because this change removes almost all lines of sight from this bin.

Our procedure for the processing of the line-of-sight probes follows from our previous papers.
For each such spectrum, we fit and subtract a 10th order polynomial to a 300 \AA\ wide section surrounding the wavelength of H$\alpha$ at the related primary galaxy redshift to remove the continuum.  
We only present and discuss H$\alpha$ fluxes in this work because the theoretical modeling of [N {\small II}] is more complicated and beyond the question we focus on here. We then measure the residual H$\alpha$ flux within a velocity window corresponding to the recessional velocity of the associated primary galaxy to capture the majority of the emission flux from the gas surrounding that galaxy. We vary the velocity window as a function of stellar mass as described  in Paper V.

We do  apply this procedure to all of the line of sight spectra. Specifically, we have a set of spectral selection criteria. First, the continuum level must be $< $ 3 $\times$ $10^{-17}$ erg cm$^{-2}$ s$^{-1}$ \AA$^{-1}$ to limit the noise introduced by the actual SDSS spectral target. Second, we require that the measured emission line flux be within 3$\sigma$ of the mean of the whole sample so as to remove  spectra of interloping strong emitters such as satellite galaxies or outer disk H{\small II} regions. We have confirmed that using the mean or median of the resulting set of flux measurements in our subsequent analysis produces consistent results. We present the results of mean flux in the following study.

We estimate the uncertainties in the mean flux values by randomly selecting half of the individual spectra in the relevant subsample, calculating the mean emission line flux, and repeating the process 1000 times to establish the distribution of measurements from which we quote the values corresponding to the 16.5, 50.0 and 83.5 percentiles. We compensate for using only half the sample in each measurement by dividing the $1\sigma$ estimated uncertainties by a factor of $\sqrt{2}$.

\subsection{H$\alpha$ Emission Dependence on the Galaxy's Stellar Mass}
\label{sec:sm}

\begin{figure}[htbp]
\begin{center}
\includegraphics[width = 0.48 \textwidth]{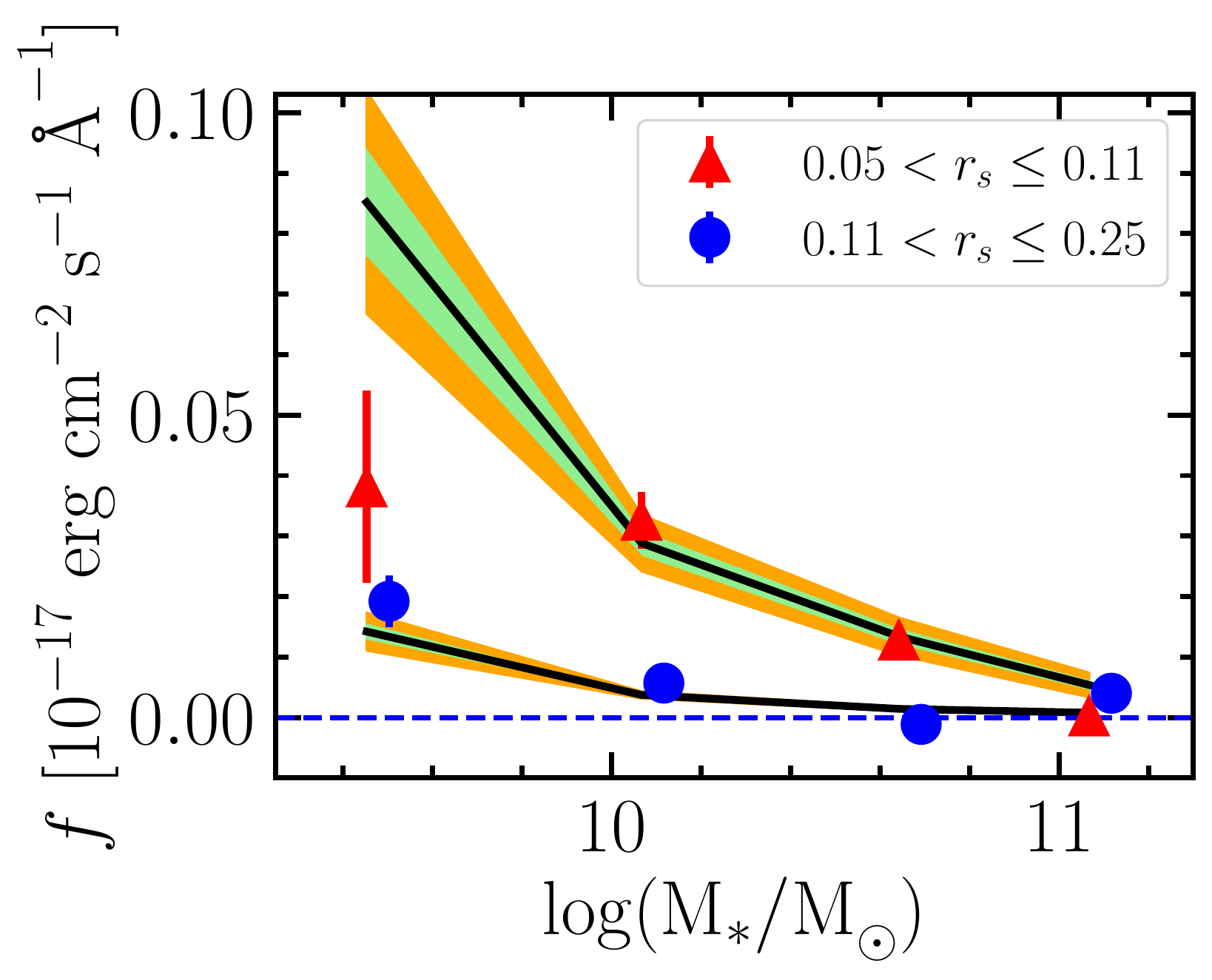}
\end{center}
\caption{Measured average H$\alpha$ flux, $f$, in two $r_s$ bins ($0.05 < r_s \le 0.11$ and $0.11 < r_s \le 0.25$) as a function of the stellar mass of the primary galaxy. For easier visualization, we apply slight horizontal offsets for the data points of the inner and outer $r_s$ bins.  The blue dashed horizontal line represents no net emission or absorption at H$\alpha$. The solid black lines represent the modeled flux, and the associated shaded regions correspond to 1$\sigma$ and 2$\sigma$ uncertainties of the model described in Sec. \ref{sec:model}.}  \label{fig:sm}
\end{figure}

We begin our analysis by re-evaluating the dependence of the H$\alpha$ flux on the primary galaxy stellar mass. 
We follow the procedure outlined
in Paper V 
and measure the H$\alpha$ flux for galaxies in different stellar mass bins. The set of primary galaxies used here, taken from SDSS DR7, is somewhat different from that used in Paper V,
which was taken from the group catalog from \cite{yang2007}. As in the previous study, we select lines-of-sight with $0.05 < r_s < 0.25$ around each primary galaxy and split the data into two bins with equal $\Delta \log r_s$. We present the H$\alpha$ flux results, as well as the average SFR which will be used in the following discussion, in Table  \ref{tab:sm} and Figure \ref{fig:sm}.  N in Tables \ref{tab:sm} and \ref{tab:environ} refers to the number of galaxies in the corresponding subsample. We use the average value of $\log$ M$_*$ and $r_s$ to reference each bin. 

In broad terms, we find that the measured H$\alpha$ flux decreases as the galaxy stellar mass increases in both $r_s$ bins, and that the flux is consistently larger in the inner $r_s$ bin.
These fluxes are quantitatively  consistent with those presented in Paper V
except for the smallest stellar mass bin. 
There are two substantive differences between this study and the previous one. We have already mentioned the first, which is that the primary galaxy samples come from slightly different sources. The second is that the line of sight spectra came from DR12 for the previous study and come from DR16 for the current study. 
We confirm that using DR12 with the current primary galaxy sample leads to insignificant differences, so we attribute the 
discrepancy to differences in the primary galaxy samples. This difference may be a statistical fluctuation or 
related to the nature of the group catalog. For example, 
low mass galaxies in the group catalog are exclusively satellites, while those in the more general SDSS population are likely to be a more heterogeneous population.

\begin{deluxetable}{cccrc}
\tablewidth{0pt}
\tablecaption{The stacked H$\alpha$  emission flux vs stellar mass and radius}
\tablehead{  \colhead{$\log({\rm M}_*$/M$_\odot$)} & \colhead{$r_s$}\tablenotemark{a} & \colhead{N} & \colhead{$f$} & \colhead{{\rm SFR}}\\
 &&&\colhead{
 [$10^{-17}$\,erg\,cm$^{-2}$\,s$^{-1}$\,\AA$^{-1}$]}&\colhead{[M$_\odot$/yr]
 }
  }
\startdata
\multirow{2}{*}{9.44} & 0.09 & 53 & $0.038 \pm  0.015$ & \multirow{2}{*}{0.70 $\pm$ 0.06} \\
& 0.19 & 611 & $0.020 \pm 0.004$ & \\ \\
\multirow{2}{*}{10.05} & 0.09 & 290 & $0.033 \pm 0.006$ & \multirow{2}{*}{1.34 $\pm$ 0.14} \\
& 0.19 & 2059 & $0.006 \pm 0.002$ & \\ \\
\multirow{2}{*}{10.64} & 0.09 & 1265 & $0.013 \pm 0.002$ & \multirow{2}{*}{1.86 $\pm$ 0.04} \\
& 0.19 & 6486 & $-0.001 \pm 0.001$ & \\ \\
\multirow{2}{*}{11.07} & 0.09 & 966 & $0.000 \pm 0.002$  & \multirow{2}{*}{1.59 $\pm$ 0.06} \\
& 0.19 & 4236 & $0.004  \pm 0.001$ & 
 \enddata
\label{tab:sm}
\tablenotetext{a}{$r_s$ is  the  ratio  between  the  physical projected separation and the virial radius of the primary galaxy, $r_s \equiv r_p / r_{\rm vir}$.}
\end{deluxetable}

\subsection{The H$\alpha$ Emission Dependence on the Galaxy's Local Environment}

Quantifying environment is always a challenge, and part of that challenge is whether to focus on a local or global measurement of environment. 
In Paper IV,
one of the environmental measurements we adopted was the projected proximity to a massive (M$_*$ $> 10^{11}$ M$_\odot$) galaxy. We justified this choice in the context of the scenario in which the CGM of a primary galaxy is removed once it enters the sphere of influence of a massive neighbor. In that work, we also explored other scenarios, such as one where the CGM of a primary galaxy is removed once it enters a group or cluster environment. Although we also found some evidence in favor of this latter scenario, the number of galaxies in rich environments is small and exploring this scenario further with the current sample proved difficult. Therefore, we proceed by further exploring only the single massive neighbor scenario and caution that environmental effects beyond that explored here may be at play.

We adjust our massive neighbor scenario slightly by lowering the required minimum massive neighbor mass to $ 10^{10.5}$ M$_\odot$ and $10^{10}$ M$_\odot$, in two separate trials, both to increase the sample size and to probe whether the physical effect continues for lower mass ``massive" neighbors. We counter
the absolute decrease in neighbor mass by requiring that every affected primary galaxy be substantially less massive (a factor of 5) than their massive neighbor. This additional criterion is set to ensure that we are still considering systems where the primary galaxy is dominated by its neighbor. We chose a factor of 5 because in our previous study (Paper VI ) 
we found that galaxy pairs with a mass ratio less than 5 and low average stellar mass ($< 10^{10.4}$ M$_\odot$) exhibit {\sl enhanced} rather than diminished H$\alpha$ emission, perhaps due to contributions from tidal material.

As before, we quantify the effect using the projected separation, $R_p$, between the affected primary and its associated massive neighbor, for potential massive neighbors that satisfy the mass criteria described above and are within recessional velocities offset of 1000 km s$^{-1}$ of the primary.
As we did before for the projected separation of lines of sight to the primary, we now scale the projected radius to the massive neighbor using the virial radius of the massive neighbor.
Also as before, we estimate the virial radius of the massive galaxy using our derived  scaling relation between the virial radius and galactic stellar mass.
We confirm that lowering the maximum velocity difference to 500 km s$^{-1}$ leads to consistent results, although it decreases the sample size by approximately 30\%. 

A series of steps is required in order to assemble our environmentally affected sample and control samples. When multiple, $i$, massive galaxies lie within $R_{p,i}/R_{\rm vir,i} < 4$ of the primary galaxy, we consider only the one with the smallest $R_{p,i}/R_{\rm vir,i}$ as the massive neighbor. To construct one control sample, we identify a sample of ``isolated" primary galaxies that have no neighbors that satisfy our massive neighbor criteria projected within 5 Mpc and lie within 1000 km s$^{-1}$ in velocity offset. For reference, the average calculated virial radius of the massive neighbors in our sample is $\sim$ 500 kpc, which implies that the typical galaxy in the ``isolated" galaxy sample is free of massive neighbors to $R_p/R_{\rm vir} \sim 10$.
Next, we require $R_p/R_{\rm vir} > 0.05$ to remove  pairs in which emission from the massive neighbor is likely to contaminate our measurement of the primary galaxy. We find that adopting twice as large a limit makes an insignificant difference in the results. 
Finally, because the primary galaxy mass distributions are different between the data sample and the control sample, we force the mass distribution of the control sample to be the same as that of the data sample at $R_p/R_{\rm vir} > 1$. We do this by randomly drawing from the larger control sample until we match the mass distribution of the data. The H$\alpha$ fluxes of the mass-matched control sample are $(0.028 \pm 0.003) \times 10^{-17}$ erg\,cm$^{-2}$\,s$^{-1}$\ \AA$^{-1}$ and $(0.011 \pm 0.001) \times 10^{-17}$ erg\,cm$^{-2}$\,s$^{-1}$\ \AA$^{-1}$, for the radial bins $0.05 < r_s \le 0.11$ and $0.11 < r_s \le 0.25$, respectively.

\begin{figure}[htbp]
\begin{center}
\includegraphics[width = 0.48 \textwidth]{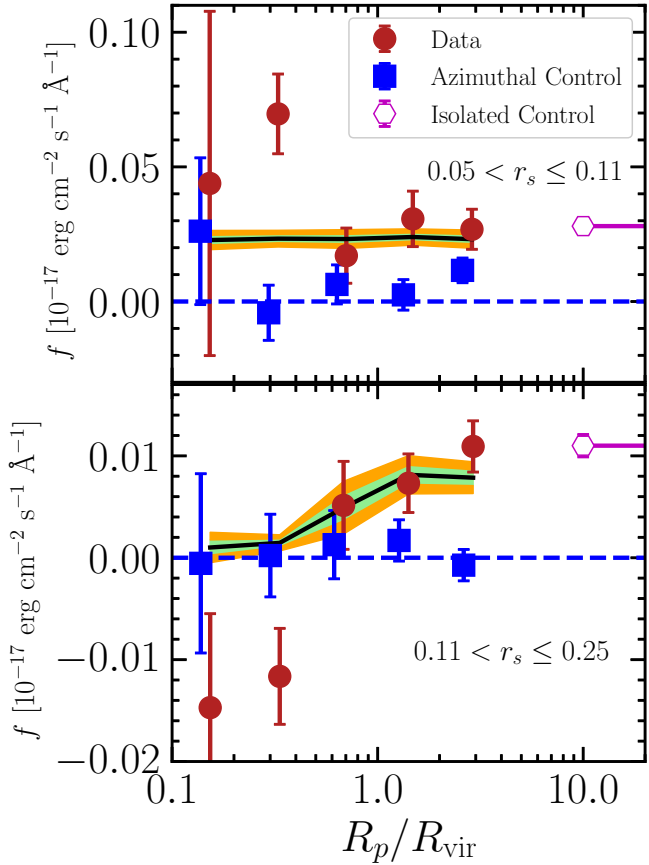}
\end{center}
\caption{Measured average H$\alpha$ flux for $0.05 < r_s \le 0.11$ (upper panel) and   $0.11 < r_s \le 0.25$ (lower panel) as a function of scaled projected distance ($R_p/R_{\rm vir}$) from the nearest massive neighboring galaxy which is at least 5 times as massive as the primary galaxy and has stellar mass greater than $10^{10}$ M$_\odot$. We also present the average H$\alpha$ fluxes for the azimuthal and isolated control samples (see text).
For easier visualization, we apply slight horizontal offsets between the data and control in the same bins. The solid black lines represents the model flux and the corresponding shaded region represent the 1 and 2$\sigma$ uncertainties of the model  described in Sec. \ref{sec:model}.
The flux for the isolated control sample is presented as a point with error bar and an attached line indicating that the value represents galaxies at large projected separations. 
Again, zero net flux is presented as a blue dashed line to guide the eye.}
\label{fig:environ}
\end{figure}

We find that the results are insensitive to our choice of the lower mass limit for the neighbors, within the range we explore, so we adopt our lowest mass limit, $10^{10}$ M$_\odot$, to maximize the sample size.
In Table \ref{tab:environ} and Figure \ref{fig:environ} we present the stacked H$\alpha$ fluxes as a function of $R_p/R_{\rm vir}$
for both the inner and outer $r_s$ bins around our primary galaxies and isolated galaxy control sample. We refer to this control sample as having $R_p/R_{\rm vir} \gtrsim 10$. We also present the mean SFRs for the primary galaxies in each subsample. Finally, we construct a second control sample in which we insert two artificial primaries at the same $R_p/R_{vir}$ as each primary, but at different angles around the massive neighbor.  We
refer to this second control sample as the ``azimuthal" control sample with $\sim$ 6000 lines of sight, in contrast to the ``isolated" control sample with $\sim$ 8000 lines of sight. 

\begin{deluxetable}{cccrc}
\tablewidth{0pt}
\tablecaption{The H$\alpha$ emission flux vs. scaled distance from nearest massive neighbor\tablenotemark{a}}
\tablehead{  \colhead{$R_p/R_{\rm vir}$}  & \colhead{$r_s$} & \colhead{N} & \colhead{$f$} & \colhead{\rm SFR}\\
& & &\colhead{[$10^{-17}$\,erg\,cm$^{-2}$\,s$^{-1}$\ \AA$^{-1}$]}&\colhead{[M$_\odot$/yr]}
  }
\startdata
\multirow{2}{*}{0.15} & 0.09 & 8 & $0.044 \pm  0.065$ & \multirow{2}{*}{0.54 $\pm$ 0.07} \\
& 0.19 & 86 & $-0.015 \pm 0.009$ & \\ \\
\multirow{2}{*}{0.33} & 0.09 & 39 &  $0.070 \pm 0.014$ & \multirow{2}{*}{0.84 $\pm$ 0.06} \\
& 0.19 & 219 & $-0.012 \pm 0.005$ & \\ \\
\multirow{2}{*}{0.68} & 0.09 & 46 & $0.017 \pm 0.011$ & \multirow{2}{*}{0.93 $\pm$ 0.05} \\
& 0.19 & 350 & $0.005 \pm 0.046$ & \\ \\
\multirow{2}{*}{1.41} & 0.09 & 67 &  $0.031 \pm 0.010$  & \multirow{2}{*}{1.14 $\pm$ 0.06} \\
& 0.19 & 625 & $0.007  \pm 0.003$ & \\ \\ 
\multirow{2}{*}{2.92} & 0.09 & 131 & $0.027 \pm 0.008$  & \multirow{2}{*}{1.56 $\pm$ 0.20} \\
& 0.19 & 1022  & $0.010  \pm 0.003$ & \\ \\ 
\multirow{2}{*}{$\gtrsim 10.0$} & 0.09 & 1137 &  $0.028 \pm 0.003$  & \multirow{2}{*}{1.17 $\pm$ 0.02} \\
& 0.19 & 6572 & $0.011  \pm 0.001$ & 
 \enddata
\label{tab:environ}
\tablenotetext{a}{The scaled distance refers to the projected separation between the primary and the massive neighbor divided by the estimated virial radius of the massive neighbor. A massive neighbor is defined to be at least 5 times as massive as the primary galaxy and have M$_* \ge 10^{10}$ M$_\odot$.}
\end{deluxetable}

We draw several conclusions from these results (Table \ref{tab:environ} and Figure \ref{fig:environ}). First, and most significantly, for the outer $r_s$ bin around the primary galaxies, the fluxes for primaries at the two smallest values of $R_p/R_{\rm vir}$ are significantly lower than for those either at larger values of $R_p/R_{\rm vir}$ or for those that are isolated. 
To confirm this impression, we calculate the nonparametric rank-order Spearman correlation and find that the null hypothesis of no correlation between flux and $R_p/R_{\rm vir}$ can be rejected at $>$ 99\% confidence in the outer $r_s$ bin. 
Second, less significantly, there is the appearance that the effect exists even beyond 
$R_p/R_{\rm vir} > 1$. Third, 
the fluxes in the inner $r_s$ bin appear to be unaffected by the presence of the massive neighbor. If anything, there may be a slight flux increase rather than decrease at small $R_p/R_{\rm vir}$, but the results for the inner $r_s$ bin have larger uncertainties due to the smaller number of lines of sight. Fourth, for both $r_s$ bins, the flux values for the outermost $R_p/R_{\rm vir}$ bin are consistent with those of the isolated control sample, indicating that we have measured out to the asymptotic value. 
Lastly,  as expected if there is no contamination from the massive neighbor, we find no mean H$\alpha$ flux and no systemic dependence of the flux with $R_p/R_{\rm vir}$ for the azimuthal control sample for both radius bins (Figure \ref{fig:environ}). 

 The one remaining puzzle is the slight systematic tendency to produce negative fluxes at small $R_p/R_{\rm vir}$ in the larger $r_s$ bin. Because we do not find the same trend in the azimuthal control sample, we exclude the possibilities that the effect is due to a systematic overestimate of the continuum in these environments or to the presence of an absorption line spectrum contribution from the massive neighbor. At this point, given its somewhat marginal statistical significance, we attribute the result to noise but note that it could also reflect stellar absorption from tidal material around the primary. 

\section{Models and Interpretation}

We now present two different modeling efforts. First, we expand our previous work of defining broad characteristics of the CGM by setting basic rules within the construct of the {\sl UniverseMachine} models and optimizing the parameter values for those rules.  Our effort here is distinguished from our previous work in that we are jointly fitting the mass and environment dependence to determine the interdependence of those, and in fitting separately to the measurements in both $r_s$ bins. The latter is important given that it is only the flux in the outer $r_s$ bin that appears affected by the nearby massive neighbor.  Second, we present a new analysis where we use the observed H$\alpha$ fluxes and our derived CGM characterization to constrain the gas mass cooling rate in the halos. In turn, when combined with the independently measured star formation rates, we will use these results to determine how much gas mass needs to be reheated or ejected in order that the central galaxy not systematically gain or lose cool gas. We then compare our measurement to independent theoretical predictions as a demonstration of where future work might lead.

\subsection{{\sl UniverseMachine} and Global Trends}
\label{sec:model}

The {\sl UniverseMachine} is a dark matter-only simulation with co-moving box length of 250 Mpc $h^{-1}$, mass resolution of $1.8 \times 10^{8} \ {\rm M}_\odot$ (2048$^3$ particles), and force resolution of $1$ kpc $h^{-1}$.  Halos were found with the \textsc{Rockstar} phase-space halo finder \citep{BehrooziRockstar}, and merger trees were generated with the \textsc{Consistent Trees} code \citep{BehrooziTrees}. The code empirically determines the dependence of galaxy star formation rates  on the host dark matter halo mass, the halo accretion rate, and redshift, and produced observables  are matched to  SDSS galaxies in stellar mass and star formation rate. In particular, galaxy colors are matched based on both stellar mass and specific SFR.

In previous studies, Papers IV and V,
we studied the dependence of the CGM on the environment and stellar mass separately. However, due to the common multiparameter correlations present among galaxies, it is possible that effects attributed to either environment or mass in those studies should rightly be ascribed to the other. Here we perform a joint modeling exercise to avoid ``double-counting" any effect. 

The central tenet of this model is the following description of the cool gas fraction that includes a simple dependence on the stellar mass of the primary galaxy, M$_*$, and environment taken in large part from our previous work: 

\begin{equation}
\label{eq:f}
\mathcal{F}_{cool}(R, {\rm M_*}) = 
  \begin{cases}
     a ({\rm M_{*,10}})^b{\left(R\over{x \cdot R_{\rm vir}}\right)} & \ \text{if} ~R < x \cdot R_{\rm vir}\\
  a({\rm M_{*,10}})^b   &  \  \text{if} ~R \ge x \cdot R_{\rm vir}
  \end{cases}
\end{equation}
where $(x, a, b)$ are free parameters, $R$ is the three dimensional distance between the primary galaxy and the massive neighboring galaxy,  $R_{\rm vir}$ is the virial radius of the massive galaxy,  which is defined to be the radius of a spherical volume within which the mean density is $\Delta_c$ times the critical density at that redshift (where $\Delta_\mathrm{c} = 18\pi^2 + 82x - 39x^2$ and $x = \Omega_m(z) - 1$), $\mathcal{F}_{cool}$ describes the fraction of the cosmological baryon fraction in the halo that is in the cool ($T \sim 10^{4-5}$K) gas phase, and $M_{*,10}$ is the stellar mass in units of $10^{10}$ M$_\odot$.  Our assumption that the baryon fraction of galaxy halos is the universal one has not yet been empirically verified. While it is observed to hold roughly for higher mass halos  \citep{Gonzalez2013,Lim2020},  it does not appear to hold universally for the lower mass halos of galaxy groups \citep[e.g.,][]{Vikhlinin2009, Gonzalez2013}. On galaxy scales we have few constraints, but the rough concurrence in the Milky Way's dynamical mass and that inferred assuming the cosmological baryon fraction indicates that the baryon fraction for large galaxies, at least for our own galaxy, does not deviate grossly from the universal value \citep{zc}. To the degree that the baryon fraction is lower than then universal value in our sample of galaxies, our results underestimate $\mathcal{F}_{cool}$.
The value of the parameter $x$ affects both the extent of the environmental and the amplitude of the effect at a given physical radius. The environmental effect is only applied to the gas in the outer $r_s$ bin, as there is no evidence for stripping of the CGM in the inner $r_s$ bin. For gas in the inner $r_s$ bin we adopt $\mathcal{F}_{cool}$ as given by the 
$R \ge x \cdot R_{\rm vir}$ condition in Equation \ref{eq:f}.
Additionally, only $\sim$ 30\% galaxies have neighbors that are at least 5 times as massive as the primary galaxy within 4$R_{\rm vir}$ of the massive neighbor. 
For those galaxies without such a massive neighbor, $\mathcal{F}_{cool}$ in 
both $r_s$ bins is defined using the $R \ge x \cdot R_{\rm vir}$ condition in Equation \ref{eq:f}. 
Our cool gas fraction functional form thus accounts for both a stellar mass and environmental dependence, which is notionally consistent with what numerical simulations find when investigating star formation/quenching \citep[e.g.,][]{Wang2018}.

We follow our earlier approach in 
Papers IV and V using  UniverseMachine \citep{Behroozi2018} catalog, and adopt a gas mass density profile for the primary galaxy consistent with that describing gas in an NFW potential \citep{NFW1996,NFW1997}. 
As we discussed in Paper II, 
the H$\alpha$ emission flux we detect comes from two contributions: the emission from the central galaxy itself and that from the spatially associated halos around the central galaxy. By constraining our measuring and modeling to within projected radii of 50 kpc, or $r_s \sim$ 0.25, from the primary galaxy, we naturally focus on the first of these components.  In Paper IV, we roughly approximated the effect of the second component of H$\alpha$ emission in those calculations by assigning half of the 
measured emission to the second component. Now, we account for both components in our models because we include the flux contributions from nearby neighbors. 
Given the simplicity of this approach, we neglect the higher order effect represented by the interaction of massive galaxies with the CGM of the other neighbors. Moreover,  the agreement  in measured fluxes between our outermost $R_p/R_{\rm vir}$ radial bin and the isolated sample, suggests that this effect is not measurable at the precision available in the current data.
Furthermore, by modeling the radial profile of the emission, in addition to the total amplitude, we are able to attempt to reproduce the measurements in our two radial bins.

\begin{figure}[htbp]
\begin{center}
\includegraphics[width = 0.48 \textwidth]{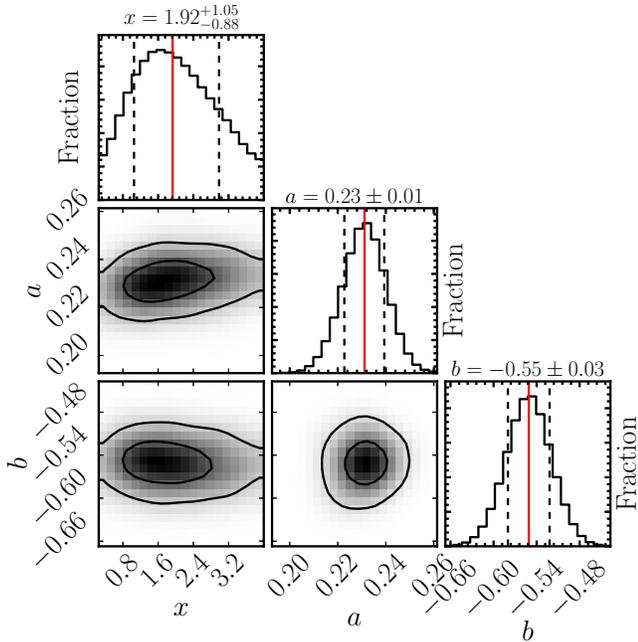}
\end{center}
\caption{Posterior distribution for the cool gas fraction in the CGM of the primary galaxy for the two radii bins of $0.05 < r_s \le 0.11$ and $0.11 < r_s \le 0.25$, as a function of distance, $R$, from the massive neighbor, and the stellar mass of the primary galaxies. The red vertical line indicates the median value of the model parameters and the shaded region highlights the 1$\sigma$ uncertainties.} 
\label{fig:modelparam}
\end{figure}

We perform a Bayesian analysis to derive confidence intervals on each of the three parameters in Eq. \ref{eq:f}. Based on the results from our previous studies we adopt uniform priors over the given ranges as follows,
\begin{equation}
0.25 < x <4.0 , \quad 0 < a < 0.5, \quad -1.0 < b < 0,
\label{eq:prior}
\end{equation}
and utilize the measured H$\alpha$ fluxes  presented in Tables \ref{tab:sm} and \ref{tab:environ}.
There is an implicit physical restriction on the model parameters that the cool gas fraction can not exceed one. Therefore, we add that constraint by setting any $\mathcal{F}_{cool} > 1$ to 1.  As standard, the posterior distribution, $p(\Theta|{\rm data})$,  is described as follows:
\begin{equation}
p(\Theta|{\rm data}) = \frac{p(\Theta) \cdot \Pi ~p({\rm data}|\Theta)}{p({\rm data})}
\end{equation}
where $\Theta$ is the parameter space $\Theta = (x, a, b)$, $p(\Theta)$ is the prior distribution described in Eq. \ref{eq:prior},  $p({\rm data}|\Theta)$ is the likelihood of the data and $p(\rm data)$ is the marginal probability for the data. 
We define the likelihood of obtaining the data given a specific model using the difference between the actual and model data, ${\rm exp[-0.5\cdot (actual - model)^2/\sigma^2}]$, where $\sigma$ is the observational  uncertainty. We use ``{\it emcee}" \cite[]{emcee} to implement a Markov chain Monte Carlo sampling of the likelihoods across the parameter space to calculate the posterior distribution. 
In Figure \ref{fig:modelparam} we display the posterior distributions of the three parameters, the median and 1$\sigma$ confidence levels (based on the 16th and 84th percentiles) for each, and the correlations between the probability distributions. 

Regarding the modeling of the environmental dependence, 
the preferred value for $x$ is $1.92^{+1.05}_{-0.88}$, demonstrating that the effects of a massive neighbor on the CGM of the primary galaxy are evident  even when the primary lies beyond  the neighbor's virial radius.
Unfortunately, the precision on $x$ is low, presumably because 
the sample is still relatively small (we only have 1447 data points for $Rp/R_{\rm vir} < 2$)  and because we only have projected distances between primaries and neighbors. Even so, the finding that $x > 1$
suggests either that the physical effect begins at these large radii or that our sample includes galaxies that have `splashed'-back, having already passed within the virial radius and currently lying outside. Various modeling and sample differences complicate the comparison to our previous result in Paper IV, but there
we estimated the effect to extend to a projected separation of $\sim$ 724 
kpc for a sample of neighbors with a mean virial radius of $\sim$ 550 kpc, or alternatively an $x$ value of $\sim$ 1.32, which is
consistent with our current estimate.

Regarding the modeling of the mass dependence, we find
$a = 0.23\pm 0.01$, which is only slightly smaller than, but consistent with, our previous estimate of $0.28^{+0.07}_{-0.06}$ in Paper V and $b = -0.55\pm 0.03$, which is significantly lower than our previous estimate of $-0.33 \pm 0.06$ in Paper V.  To explore the cause of this 
marked change, we model the stellar mass dependence and the environmental effect separately, as done in our previous studies. Our new best fit for $b$ is $-0.55 \pm 0.03$, demonstrating that the difference with our previous result does not lie with the new joint analysis of mass and environment. 

We ascribe the significant difference in $b$ to two factors. First, there is a statistically significant difference in the measured flux for the lowest mass galaxies in the innermost $r_s$ bin between  the two studies. The flux was  lower in the previous study and this might be related  to the selection of group galaxies in that previous work, most of which at these low masses would be satellites and might differ from a more general sample of low mass galaxies. Second, the basic power-law model we use is manifestly overpredicting the measured flux for these low mass galaxies at the smallest separations (Figure \ref{fig:sm}), suggesting that this model may not be an accurate representation and that the dependence of $\mathcal{F}_{cool}$ with mass may not be quite as steep as indicated by our new estimate of $b$.

We show in Figures \ref{fig:sm} and \ref{fig:environ} that the most likely model is also one that reasonably reproduces the observational data. With regard to the mass dependence of $\mathcal{F}_{cool}$, we find a strong decline in the cool gas fraction with increasing stellar, and presumably total, mass is needed.

With regards to the environmental effects, we interpret the model as indicative of the removal of the outer portions of the CGM of the primary galaxy by the massive neighbor \citep{Putman2021}. The effect is significant even when the primary galaxy falls in the potential well within 2$\times R_{\rm vir}$ of the massive neighbor, and  most  severe when the primary galaxy lies within the virial radius of the massive neighbor.  The model suggests that satellites rapidly lose some of their gaseous haloes, or that their hot gaseous halos quickly become unable to cool after reaching the virial radius of a larger halo. The apparent onset of the effect at radii larger than the virial radius could reflect the nature of the splashback radius \citep[for discussion see][]{more2015}, where galaxies that have fallen inside of the virial radius return to larger radii as they execute their orbits, or on the affected nature of mass accretion onto halos at these radii \citep{Behroozi2014}. The significant removal of the gas but only small decrease in the SFR might indicate the delayed quenching of those satellites in rich environments \citep{Wetzel2013}.

\subsubsection{An Aside on H$\alpha$ Emission and Galaxy's Color}

As is always the case for galaxy studies, a variety of properties are related to each other and this complicates the analysis and interpretation. We just measured that H$\alpha$ flux from the CGM is inversely related to the stellar mass of the galaxy, but stellar mass is related to a number of other galaxy properties, such as color and morphology. Is it possible that a relationship between H$\alpha$ flux and galaxy color, which is a parameter that is tabulated for our sample, provides an easy-to-access additional constraint on our modeling? 

\begin{figure}[htbp]
\begin{center}
\includegraphics[width = 0.48 \textwidth]{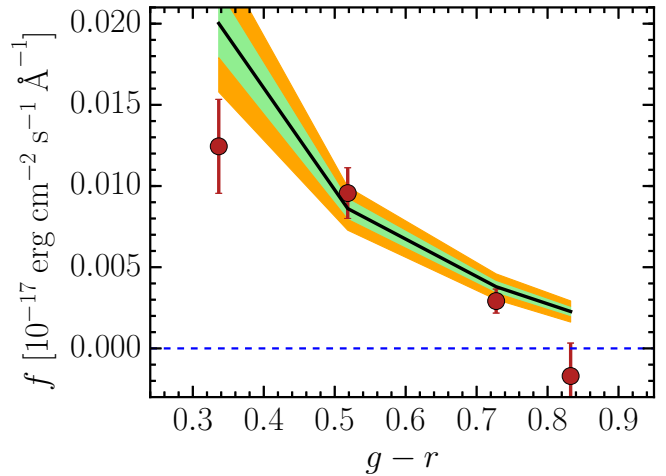}
\end{center}
\caption{Measured average H$\alpha$ flux for $0.05 < r_s  < 0.25$ as a function of  $g-r$. The solid black line represents the model flux and the corresponding shaded regions represent to 1 and 2$\sigma$ uncertainties of the model  described in Sec. \ref{sec:model}. Again, zero net flux is presented as a blue dashed line to guide the eye. } 
\label{fig:colorsersic}
\end{figure}

At a given stellar mass, one might expect blue, star-forming galaxies to have a richer gas reservoir. 
Indeed, we do find that the bluer galaxies have larger H$\alpha$ emission, as shown in Figure \ref{fig:colorsersic} and that color is highly correlated to the galaxy stellar mass (for $g-r$ bins centered on 0.34, 0.52, 0.73, 0.83, the corresponding mean stellar masses are $10^{9.8} \rm M_\odot$, $10^{10.34} \rm M_\odot$, $10^{10.76} \rm M_\odot$, $10^{10.91} \rm M_\odot$, respectively). 

We now use the UniverseMachine models to see if there is potential additional information in the relationship between H$\alpha$  flux and galaxy color. Galaxy color is included in the output of our previous modeling, so we now compare the model predictions to the data in Figure \ref{fig:colorsersic}. We find good agreement, which is at a minimum a reassuring consistency check and
suggests that galaxy color is highly degenerate with galaxy stellar mass. The result does not inform us as to whether stellar mass or color is the primary driver of the observed trend in flux, only that  ascribing it to one or the other is likely to produce similar results. Given the expected rise in the virial temperature of  galaxy halos with halo mass, which is presumably tracked by stellar mass, we prefer to base our simple modeling of $\mathcal{F}_{cool}$ on stellar mass than on galaxy color.

\subsection{Modeling the Cooling Mass Rate and Inferring the Mass Loading Factor} 
\label{sec:cooling rate}

We do not have a quantitative, complete,  empirically-validated understanding of how star forming galaxies sustain their star formation. On one hand, ongoing star formation in a large fraction of galaxies demonstrates that star formation lasts at least $\sim$ 10 billion years in those systems, but on the other hand,  the depletion time of the molecular gas for low-redshift normal galaxies is in the range of $1-2$ Gyr \citep{Saintonge2011,Tacconi2018}.
The natural solution is that the galaxy disk is being refueled and that the source of that fuel is, by consensus, the CGM \citep{borthakur2015,CGM2017}.   

Our measurement of the H$\alpha$ flux coming from the CGM provides an opportunity to measure the rate at which the CGM is radiatively losing energy, and hence cooling.
Because the radiative energy loss must lead to the equivalent loss of thermal energy in the gas, we will estimate the rate at which CGM gas is cooling and, presumably, sinking into the central disk. 
A complication that we address below is that
H$\alpha$ is only one of the emission lines through which the gas is cooling.
Assuming that the disk maintains a certain gas mass on average over some current modest time interval, the excess of the mass inflow rate over the star formation rate is the amount of gas that is reheated or ejected. This is the mass loading factor of whatever feedback mechanism is at work regulating star formation. 

Once we have calculated the total radiative losses across the entire wavelength range (\S\ref{sec:cloudy}), we will set that equal to the thermal energy loss in the gas. Assuming an ideal gas, the energy radiated per particle as it changes in temperature by $\Delta T$ is 
\begin{equation}
\Delta E = \frac{3}{2} k\Delta T.
\label{eq:E}
\end{equation}
In our calculation, we divide the entire temperature range into a few bins and $\Delta T$ is the change in temperature of the gas parcel that is considered to contribute to the measured H$\alpha$ flux. For a mean particle mass of $\mu m_p$, where $m_p$ is the proton mass, the rate at which mass is cooling (the mass cooling rate), $\dot{m}$ is:

\begin{equation}
\dot{m} = \dot{E}/\Delta E \times \mu m_p,
\label{eq:n}
\end{equation}
where $\dot{E}$ is the radiated energy rate for the whole halo and is calculated as  follows:

\begin{equation}
\dot{E} =  R_{\rm fiber} \times f_{\rm H \alpha} \times C_{\rm Cl}
\label{eq:L}
\end{equation}

\noindent
where $R_{\rm fiber}$
is the ratio of the flux contained within the fiber aperture to that contributed by the full halo, $f_{\rm H \alpha}$ is the H$\alpha$ flux measurements described above in units of erg cm$^{-2}$ s$^{-1}$, and $C_{\rm Cl}$ is the unitless correction factor derived from \cloudy modeling (\S\ref{sec:cloudy}) that converts the H$\alpha$ radiative energy loss to that across all emission lines.  $R_{\rm fiber}$ is described as $
\pi (D_{\rm A}  r_{\rm fiber})^2  \times  C_{\rm G}$, 
where $D_{\rm A}$ is the angular diameter distance at the redshift of our primary galaxy sample, which on average is approximately 0.1, $r_{\rm fiber}$ is the eBOSS angular fiber size, and $C_{\rm G}$ is a unitless geometric correction factor that we describe in \S\ref{sec:corr}. 

\subsubsection{Determining $C_{\rm Cl}$}
\label{sec:cloudy}

We model the dependence of the emission spectrum from the CGM  on the ionization state ($\log {\rm U}$), the metallicity ([X/H]), the gas temperature (T),   the total hydrogen column density ($\log {\rm N_{H}}$) and hydrogen density ($\rho_{\rm H}$) using \cloudy \citep{Cloudy98,Cloudy13,Cloudy17} version 17.02. For the ionization spectrum we adopt either the  \cite{Haardt2001} background radiation field from quasars and galaxies (HM 2001) or the Kurucz79 O star spectrum corresponding to an effective temperature of 30,000\,K of the star \citep{Kurucz1979}. We assume that the gas is in ionization equilibrium.

We find that the results are relatively insensitive to the choice of ionization source spectrum (resulting in $< 10\%$ changes to the mass cooling rate estimated below) and so we only present results from calculations using the O star spectrum as the ionization source.  

One complication that arises when converting the H$\alpha$ flux into an energy loss rate is that the halo gas emitting H$\alpha$ is not at a single temperature, but rather consists of gas at a large range of temperatures, approximately $10^{4-5}$ K  \citep[e.g.,][]{Schure2009,Lykins2013}, all of which can contribute to the observed flux. We present results from a calculation to illustrate how the flux and inferred cooling rate depend on the gas temperature T. 
For this exercise we set the other parameters to those inferred by  \cite{Werk2014} and  \cite{prochaska2017}: ionization parameter $\log {\rm U}$ = $-$3.0, metallicity [X/H] = $-$0.75, total hydrogen column density $\log ({\rm N_{H}}$/cm$^{-2}$)  = 19.5, and hydrogen density $\rho_{\rm H}$ = 10$^{-3.0}$ cm$^{-3}$. 
We do not include dust in the \cloudy modelling. The interplay of total hydrogen column density and the hydrogen density determines the geometry of the clouds, in other words, it sets the total length of the gas cloud along the line of sight. We vary the gas temperature T from 10$^{3.5}$\,K to 10$^6$\,K.
The selected temperature range roughly covers all the different phases for the halo gas in the CGM for a Milky Way like galaxy that will emit H$\alpha$.

\begin{figure}[htbp]
\begin{center}
\includegraphics[width = 0.48 \textwidth]{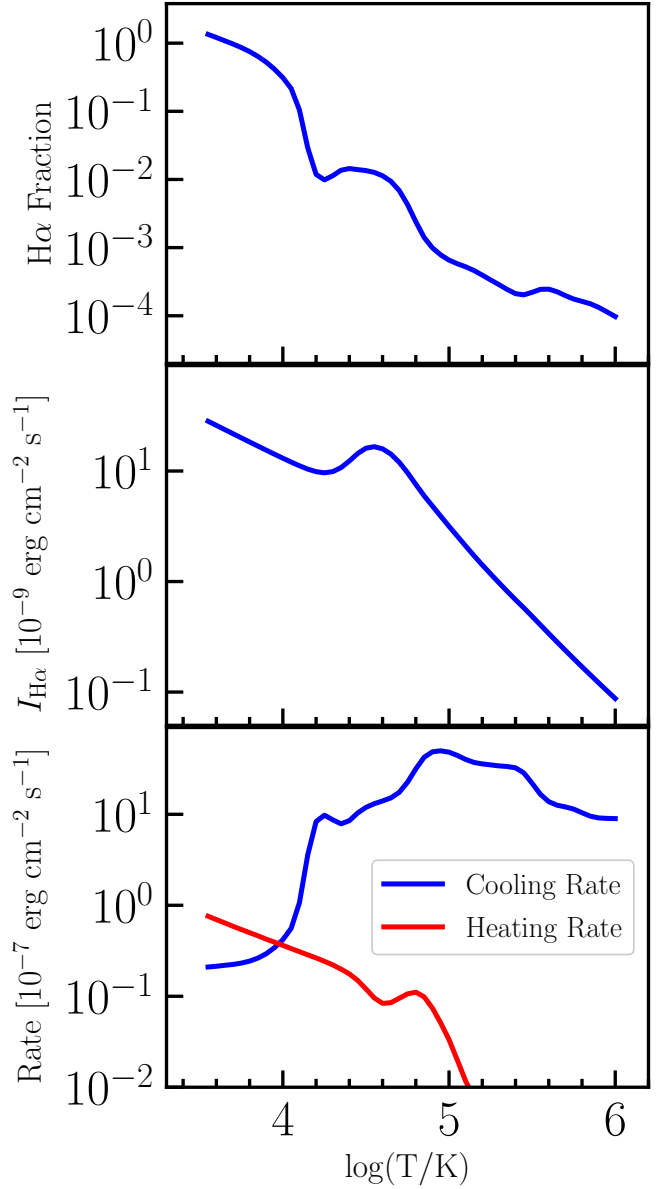}
\end{center}
\caption{The fraction of the total cooling rate attributed to the H$\alpha$ emission line (top), the H$\alpha$ line intensity (middle), and the total cooling/heating rate (bottom) as functions of the gas temperature as calculated by \cloudy for  $\log \rm U = -3.0$, [X/H] = $-0.75, \log$ (N$_{\rm H}$/cm$^{-2}$) = 19.5, and $\rho_{\rm H}$ = 10$^{-3.0}$ cm$^{-3}$.}
\label{fig:temp}
\end{figure}

The fraction of all the cooling radiation that comes from H$\alpha$ emission, what we refer to as the H$\alpha$ fraction, the H$\alpha$ radiation intensity, and the total cooling/heating rate are shown as functions of temperature in Figure \ref{fig:temp}. 
At higher ionization ($\log U > -3.5$) and temperatures $< 10^4$ K the cooling rate drops by almost two orders of magnitude and the heating rate exceeds the cooling rate, suggesting that this simple model is missing cooling channels, such as molecular lines. Therefore, we set the lower boundary of the temperature range we will consider at 10$^{4}$ K. At the high temperature end (T $ > 10^5$ K), the H$\alpha$ intensity and the H$\alpha$ fraction drops rapidly, so we also consider the contribution of this gas to H$\alpha$ negligible. At lower ionization ($\log U < -3.5$), the heating rate is orders of magnitude lower than the cooling rate at all temperature. Although we have limited the temperature range of the gas that is relevant to our calculation, the range remains large, $10^{4-5}$ K, and 
H$\alpha$ fraction and intensity vary significantly across this range, suggesting  the estimate of the distribution of gas across this range of temperatures is fundamental to understand the origin of H$\alpha$ properties.  

Unfortunately, this information is currently unavailable to the required accuracy and, in order to make progress, we need to introduce additional assumptions. Because the cooling rates vary with temperature, there are naturally temperature regimes that the gas transitions through faster and slower as it cools. To avoid gas piling up
at certain temperatures, we enforce
a continuous cooling mass flow through the  temperature regimes by scaling the gas mass at the different temperatures inversely by its corresponding  mass cooling rate.  As expected, this implies that most of the gas ends up at temperatures near $10^4$ K, where the cooling rate falls steeply. This approach also means that if one calculates the mass cooling rate at any single temperature, that value is the mass cooling rate at all temperatures.
 
In addition, we also require the integrated H$\alpha$ flux in the temperature range of $10^4 - 10^5$ K to match the observational values shown in Table \ref{tab:sm} and \ref{tab:environ}. After scaling, the integrated H$\alpha$ intensity over the temperature range of $10^4 - 10^5$ K  is $0.81\times 10^{-9}$ erg cm$^{-2}$ s$^{-1}$. The aperture size of eBOSS is 2 arcsec, which gives a conversion factor from intensity to flux of $9.41\times 10^{-11}$, then the integrated H$\alpha$ flux over the SDSS fiber size in the the temperature range of $10^4 - 10^5$ K is $0.06\times 10^{-17}$ erg cm$^{-2}$ s$^{-1}$.  The  mass cooling rate corresponding to this H$\alpha$ flux is estimated to be 0.15\,M$_\odot$/yr before geometric correction.

This approach allows us to estimate the mass cooling rate from the observed H$\alpha$ fluxes, building up on the derived scaling relations between the \cloudy H$\alpha$ flux and the \cloudy mass cooling rate. Nonetheless, this result may be sensitive to our assumptions in the \cloudy modelling, as we discuss below.

\subsubsection{Determining $C_{\rm G}$}
\label{sec:corr}

To determine the geometric correction factor, ${C_G}$, we 
take a two step process.
First, we correct for the fraction of the projected area covered by the eBOSS fiber within the radial range probed by our study ($0.05 < r_s \le 0.25$).
Because we average results from thousands of fibers scattered throughout this area with no correlation to the primary galaxy, the mean flux in a fiber aperture accurately represents a
sampling of the flux within the full area, modulo the ratio of the fiber aperture to the full aperture. As such, 
we simply correct by the ratio of the two areas $(A_{r_s<0.25}/A_{\rm fiber})$.  The correction factor varies for different galaxies because the physical radii are different even for the same scale radius range of $0.05 < r_s < 0.25$. Then the mean correction factor is 14.3, 36.5, 128.9, 682.5 for the four stellar mass bins described above. 
Second, we consider the correction for the ratio of the full halo to that which the $0.05 < r_s \le 0.25$ annulus represents,  which we define as $\mathcal{F}_{corr}$. 
We find in our model that although there is nearly as much baryonic mass in the gaseous halo outside $r_s = 0.25$ as inside, there is a negligible contribution to the H$\alpha$ flux (see Figure 7 in \citep{zhang2018a}). Therefore, $\mathcal{F}_{corr}$ is approximately equal to 1.
Our calculation for $C_G$ is 

\begin{equation}
 C_G = \frac{A_{r_s < 0.25}}{A_{\rm fiber}} \times \mathcal{F}_{corr},
\label{eq:CG}
\end{equation}

\noindent 
where we set $\mathcal{F}_{corr} = 1$.

\subsubsection{Uncertainties in the Derived Mass Cooling Rate}

To develop intuition regarding the sensitivity of our results to various parameter choices, 
we investigate the dependence of the total mass cooling rate on the adopted ionization parameter ($U$), the metallicity ([X/H]),  and the hydrogen density ($\rho_{\rm H}$), while keeping the column density fixed.  \cite{prochaska2017}  presented a radial distribution of hydrogen column density measurements in $L_*$ galaxies, roughly ranging from $\log {\rm (N_{H}/{\rm cm}^{-2})} = 20$ to $\log {\rm (N_{H}/{\rm cm}^{-2})} = 19$ for projected radius between 10 and 50 kpc, respectively, so we adopt $\log {\rm (N_{H}/{\rm cm}^{-2})} = 19.5$.
For the other parameters we allow that:

\begin{enumerate}
\item $ \log {\rm U} $ vary from $-$6 to $-$1 (our preferred parameter range is from $-$4 to $-$2),

\item $[$X/H] vary from $-$2.5 to 0.5 (our preferred parameter range is from $-$1.5 to 0), and

\item $\rho_{\rm H}$ vary from 10$^{-2.0}$ cm$^{-3}$  to 10$^{-4.0}$ cm$^{-3}$ (our preferred parameter range is from $10^{-2.5}$ to $10^{-3.5}$ cm$^{-3}$). 
\end{enumerate}

\noindent
Our preferred parameter ranges are  based on the constraints from  absorption line measurements \citep{Werk2014, prochaska2017} using QSO spectra from the COS-Halos Survey \citep{Werk2013}.  Given the fixed column density, the range of hydrogen densities we probe corresponds to a column length of the simulated halo between 10 kpc and 300 kpc. 

\paragraph{Dependence on the Ionization Parameter}

The ionization parameter, which measures the number of ionizing photons per atom, affects the calculated heating and cooling rates, and therefore the resulting H$\alpha$ flux. Here, we fix the other parameters to be [X/H] = $-$0.75 and $\rho_{\rm H}$ = 10$^{-3.0}$ cm$^{-3}$, and vary only $\log \rm U$ and the temperature. As described above, for each ionization parameter, we enforce continuity in the mass gas flow rate at different temperatures (\S \ref{sec:cloudy}) and then estimate the mass cooling rate. 

\begin{figure}[htbp]
\begin{center}
\includegraphics[width = 0.48 \textwidth]{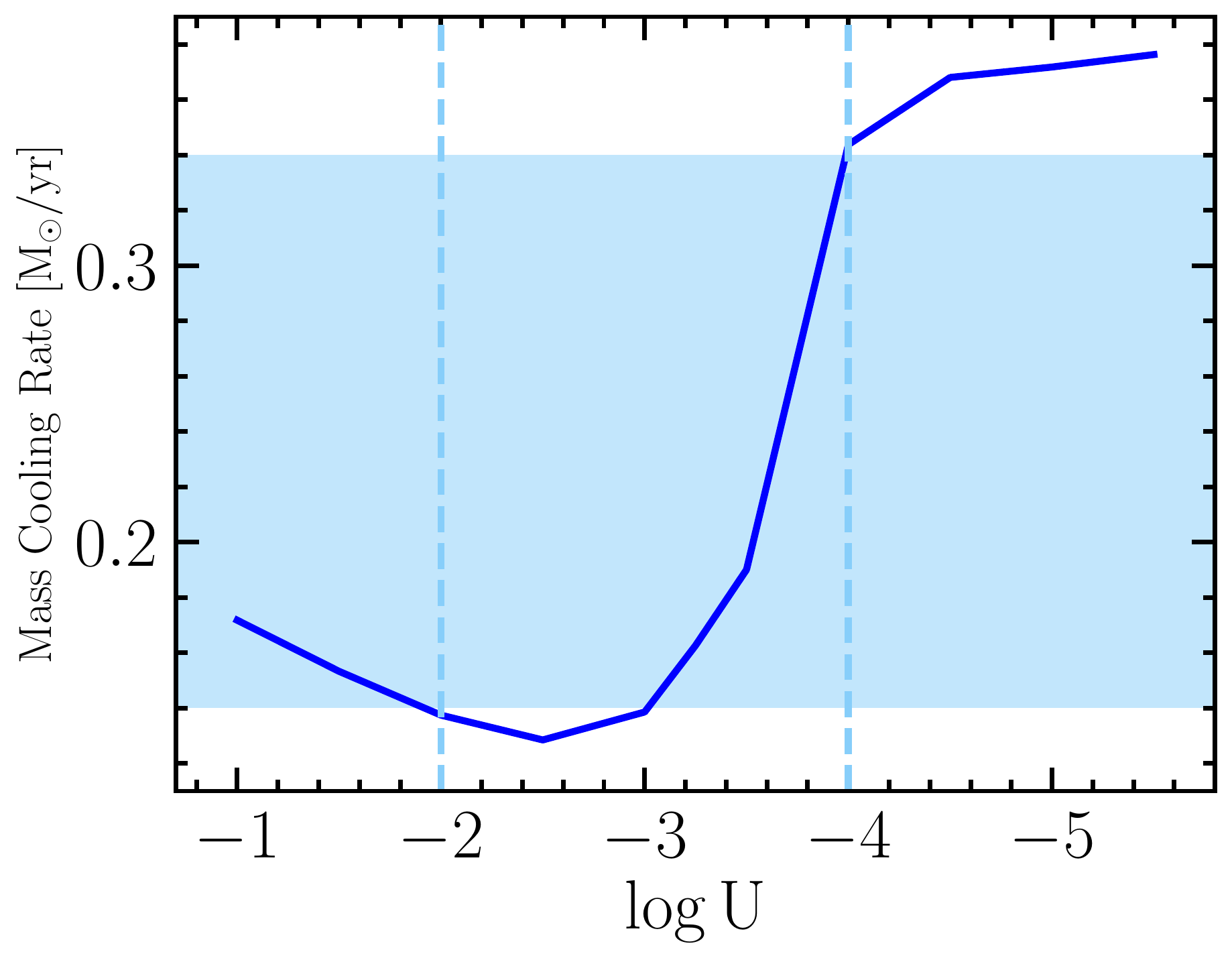}
\end{center}
\caption{The  mass cooling rate as a function of the ionization parameter, $\log \rm U$, for fixed [X/H] = $-0.75, \log ({\rm N}_{\rm H}$/cm$^{-2}$) = 19.5, and $\rho_{\rm H}$ = 10$^{-3.0}$ cm$^{-3}$. What we consider to be the plausible range of $\log {\rm U}$, based on \cite{Werk2014}, is identified by vertical dashed lines and the corresponding range of plausible mass cooling rates as the shaded region. } 
\label{fig:ion}
\end{figure}

The result is shown in Figure \ref{fig:ion}. 
We find that changes of about a factor of three in the estimated mass cooling rate are possible, although the change is rather sudden at about $\log {\rm U} \sim -3.5$. At our preferred value of $\log {\rm U} = -3$ we expect to be calculating the comparably lowest likely values of the mass cooling rate and plausible changes in $\log {\rm U}$ should only produce higher estimates of the mass cooling rate.

\paragraph{Dependence on Metallicity}

Metals contribute significantly to the total cooling in the temperature range of $10^{4-5}$ K \citep[e.g.,][]{Schure2009,Lykins2013}. Here we investigate how the adopted gas metallicity affects the inferred mass cooling rate. We adopt $\log {\rm U} = -3.0$, and $\rho_{\rm H}$ = 10$^{-3.0}$ cm$^{-3}$, and vary the gas metallicity and the temperature. Again, we enforce gas cooling continuity across temperature ranges and estimate the mass cooling   rate, as described in \S\ref{sec:cloudy}. 

\begin{figure}[htbp]
\begin{center}
\includegraphics[width = 0.48 \textwidth]{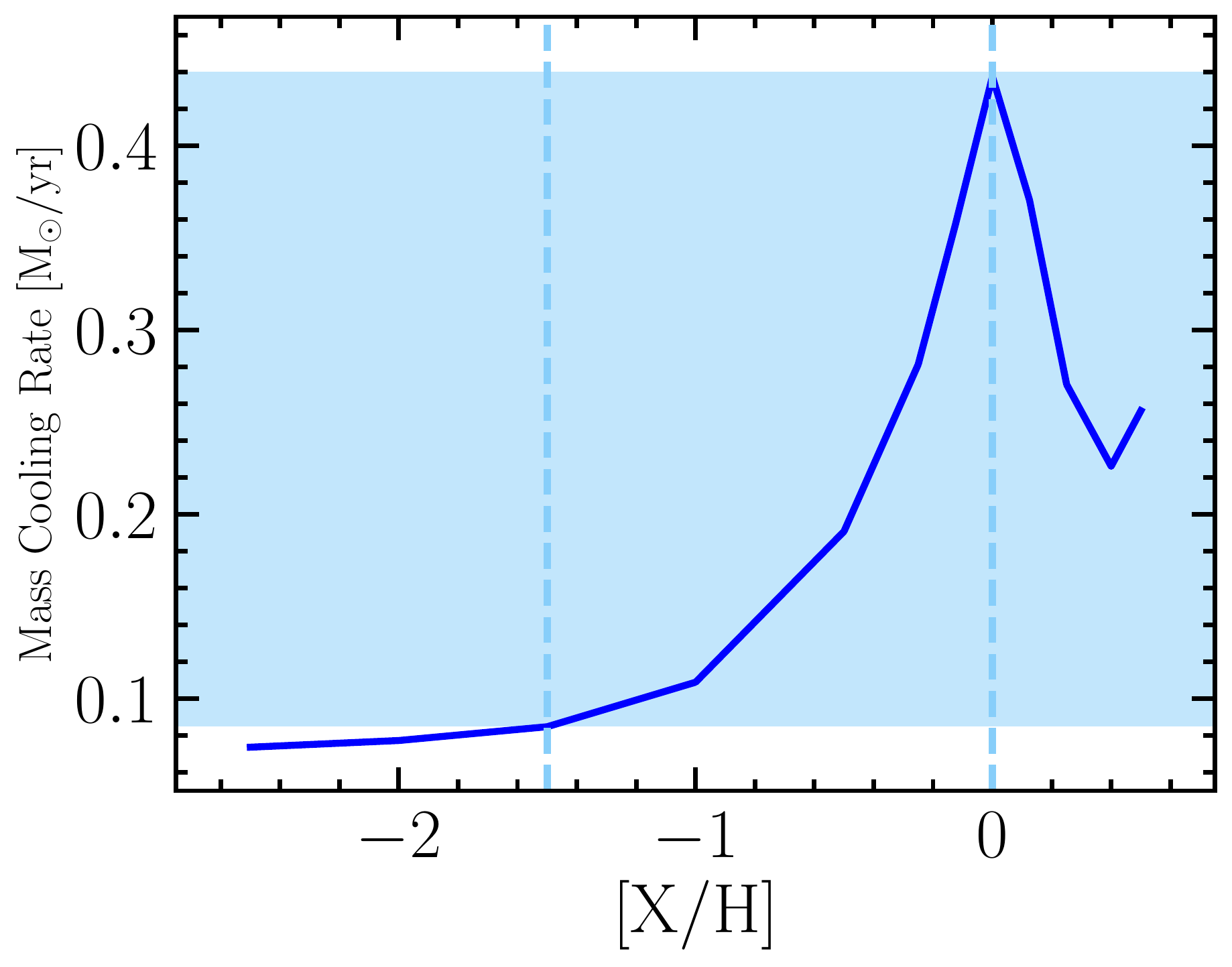}
\end{center}
\caption{The mass cooling rate as a function of the metallicity, [X/H], for fixed $\log \rm U = -3.0$, $\log ({\rm N}_{\rm H}$/cm$^{-2}$) = 19.5 and $\rho_{\rm H}$ = 10$^{-3.0}$ cm$^{-3}$.
What we consider to be the plausible range of [X/H], based on \cite{Werk2014} and \cite{prochaska2017}, 
is identified by vertical dashed lines and the corresponding range of plausible mass cooling rates as the shaded region.} \label{fig:metal}
\end{figure}

The result is shown in Figure \ref{fig:metal}. As expected, the  mass cooling rate increases  as the metallicity increases for metallicity below solar metallicity, resulting in the inferred mass cooling rate increasing significantly as we adopt a higher metallicity of the gas.
At our preferred value of [X/H] = $-$0.75, we are near the lower limit of the inferred mass cooling rates. If the CGM has solar metallicity we would be underestimating the mass cooling rate by about a factor of 2 for our standard assumed [X/H], but it is highly unlikely that the CGM reaches solar metallicity (\cite{prochaska2017} find that the typical CGM metallicity is $\sim -0.5$). For more plausible errors in the metallicity, the corresponding errors are  significantly smaller. 

\paragraph{Dependence on the Density}

\begin{figure}[htbp]
\begin{center}
\includegraphics[width = 0.48\textwidth]{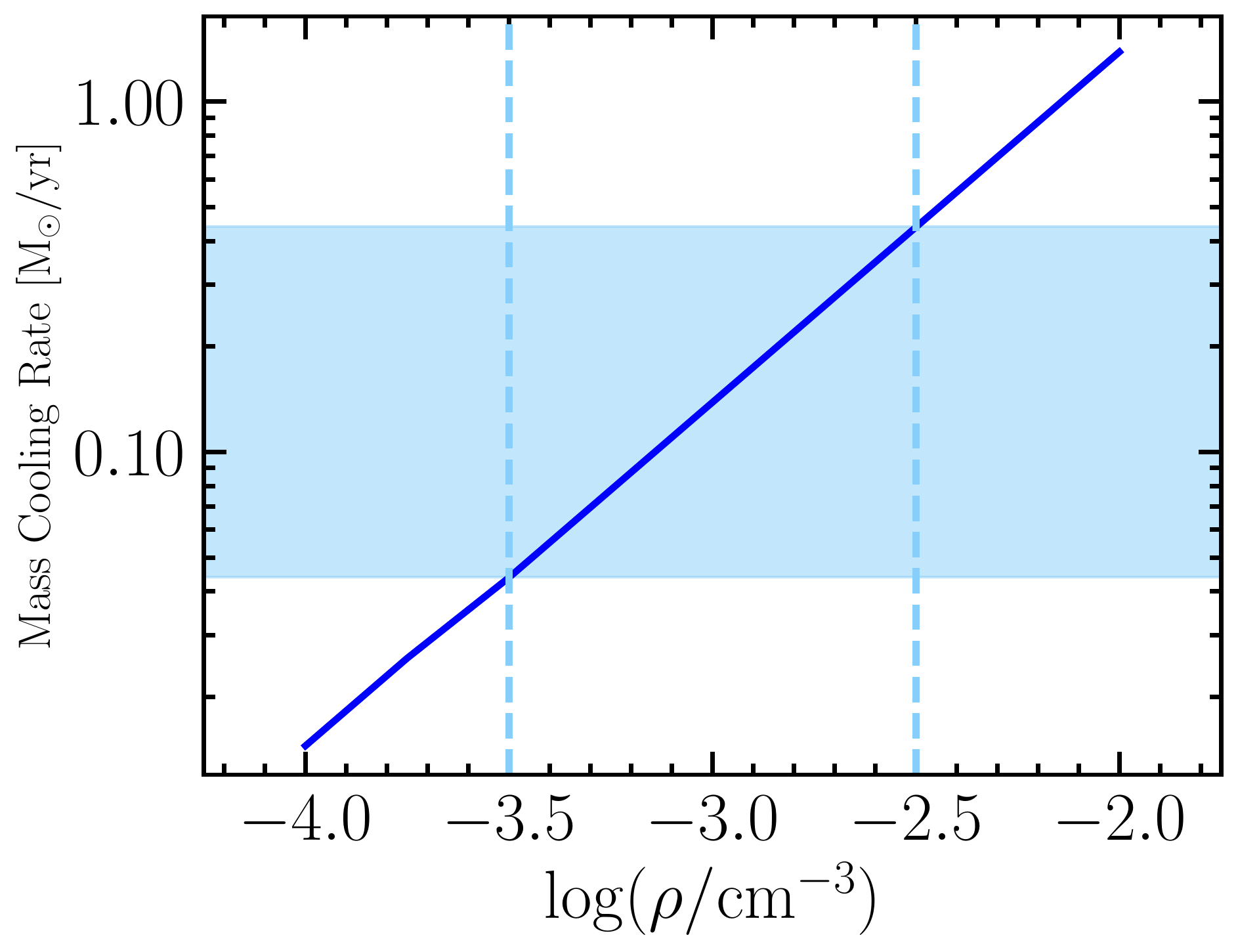}
\end{center}
\caption{The   mass cooling rate dependence on the gas density  ($\rho_{\rm H}$) for fixed $\log \rm U = -3.0$, [X/H] = $-0.75$ and $\log ({\rm N}_{\rm H}$/cm$^{-2}$) = 19.5. What we consider to be the plausible range of $\rho_{\rm H}$, based on \cite{Werk2014} and \cite{prochaska2017}, 
is identified by vertical dashed lines and the corresponding range of plausible mass cooling rates as the shaded region.}
\label{fig:rho}
\end{figure}

Both the cooling rate and the H$\alpha$ flux are proportional to the gas density squared. Because we have fixed the total hydrogen column density, our inferred cooling rates and H$\alpha$ fluxes are proportional to the gas density, not the density squared. In the \cloudy calculation, we have adopted a uniform gas density as a simple approximation to the halo gas density profile, which roughly follows a power law according to \cite{Werk2014} in other simulations. 
Adopting $\log {\rm U} = -3.0$ and [X/H] = $-0.75$, we obtain the results shown in Figure \ref{fig:rho}. 
As expected, the mass cooling rate is proportional to the adopted gas density. 

We close this section by noting that the model we present is a highly oversimplified version of the CGM. Highlighting a few specific shortcomings, we note that 1) the density we model with  \cloudy  is independent of the galactocentric distance, whereas we know the true density of the galaxy halo is decreasing significantly with radius, 2)  we do not consider density variations within a given column (i.e. clumpiness), and 3)  we only consider the variation of $\mathcal{F}_{cool}$ with galaxy mass and ignore other potential mass dependencies, such as one between CGM metallicity and mass. The model we present is best at establishing broad behavior and testing our conceptual understanding rather than at providing a quantitative understanding. Ultimately, the data presented here will provide the strongest constraints when compared with forward-modeling of physically-motivated, high-resolution CGM simulations.

\subsection{The Mass Cooling  Rate}
\label{sec:mcr}

\subsubsection{Empirical Results}

We integrate the flux within the radial range of $0.05 < r_s \le 0.25$ over the velocity window described in \S\ref{sec:dataAna} and then apply the appropriate geometric correction described in Eq. \ref{eq:CG} to obtain the CGM H$\alpha$ emission within the virial radii of galaxies in the different stellar mass bins. 
We also calculate the CGM H$\alpha$ flux using our fitted model for $\mathcal{F}_{cool}$ in Eq. \ref{eq:f} and the best fit parameter values for the assumed $T =12,000$ K.  We will present both results.

To infer a mass cooling rate (MCR) from those emission fluxes, we adopt our preferred  parameter set ($\log \rm U = -3.0$, [X/H] = $-$0.75, $\log ({\rm N}_{\rm H}$/cm$^{-2}$) = 19.5, and $\rho_{\rm H} = 10^{-3.0}$ cm$^{-3}$ \citep[from][]{Werk2014, prochaska2017}. From Figures \ref{fig:ion} to \ref{fig:rho}, we know that the mass cooling rate depends significantly on the adopted \cloudy model parameters. To estimate the effect of these uncertainties, we randomly  uniformly distribute 10,000 points in the 3-D parameter space within the range of $(-4 < \log {\rm U} < -2,  -1.5 < {\rm [X/H]} < 0,  10^{-3.5} < \rm \rho_H/cm^{-3} < 10^{-2.5})$.  We follow the steps discussed in Sec. \ref{sec:cloudy} to calculate the mass cooling rate from the H$\alpha$ flux derived from our model for $\mathcal{F}_{cool}$ using  \cloudy. We adopt the standard deviation among those 10,000 estimates of the mass cooling rate as the 1$\sigma$ uncertainty of the modeled mass cooling rate.

\begin{figure}[htbp]
\begin{center}
\includegraphics[width = 0.48 \textwidth]{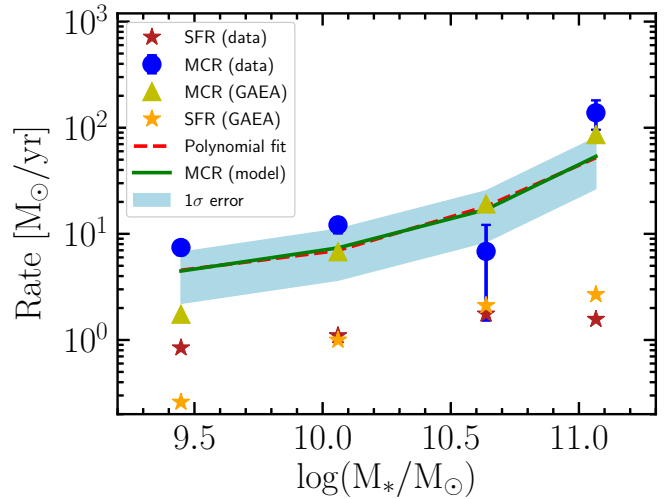}
\end{center}
\caption{The total mass cooling rates, MCRs, and star formation rate, SFRs, of the primary galaxies as a function of stellar mass.
The error bars on the blue circles (the cooling rates calculated from our H$\alpha$ flux measurements) reflect only the flux measurement uncertainties and do not reflect any uncertainty in the adopted \cloudy parameters. In contrast, the mass cooling rates calculated from our fitted model for $\mathcal{F}_{cool}$, the solid green line, include the systematic modeling uncertainties (shaded region). We also plot the mass cooling rates predicted by the GAEA simulation  \citep{xie2020} as triangles.
The SFRs both empirical (red stars) and as predicted in the GAEA realizations (yellow stars) provide benchmark comparisons. Finally, as the red dashed line we also plot our polynomial fit to our model (Equation \ref{eq:mcr}).} 
\label{fig:smcooling}
\end{figure}

We present our results for the mass cooling rates
in Figure \ref{fig:smcooling} and compare those rates to the current SFR as a function of galaxy stellar mass. As expected, the mass cooling rates are much higher across the board than the SFR of the central galaxies. We draw several conclusions from this comparison: 1) for our preferred parameters  the mass cooling rate is well above what is needed to maintain the current rate of star formation, thereby addressing the gas consumption problem, 2) this  estimated mass cooling rate significantly exceeds what is needed and therefore supports the theoretical understanding that the conversion of cooling gas into stars is inefficient and that a large fraction of the gas must be reheated or ejected, and 3) the strength of this ``feedback" is the largest for the galaxies with the largest stellar masses, but otherwise appears roughly constant as a function of stellar mass across this limited mass range. The approximate linearity in log-log indicates that  the relation between $\log \rm MCR$ and $\log \rm M_{*,10}$, where $M_{*,10}$ is the stellar mass in units of $10^{10}$ M$_\odot$, can be approximated by a second order polynomial function as follows:
\begin{equation}
\log {\rm MCR} = 0.81 + 0.48 \log {\rm M_{*,10} + 0.35 \log^2 M_{*,10}}
\label{eq:mcr}
\end{equation}

When drawing quantitative conclusions from our results, such as inferring the fraction of the gas that needs to be reheated or ejected to maintain the SFR at a steady state, one must bear in mind the large systematic uncertainties that exist. The error bars as plotted on the individual data points  reflect only the uncertainties due to the measurement errors on the H$\alpha$ fluxes. The uncertainties plotted on the model predictions reflect systematic uncertainties in the adopted parameters. In either case, our previous qualitative conclusions drawn from Figure \ref{fig:smcooling} hold, but these uncertainties, particularly the systematic ones, obviously significantly affect a quantitative estimate of the ratio of the cooling mass that must be reheated or ejected relative to that resulting in new stars. 

\subsubsection{Comparison to Previous Theoretical  Results}

Given the importance of gas cooling to galaxy evolution, there is a large set of existing literature examining this topic. Of particular relevance to our results, there is theoretical work estimating the gas cooling flow. For example,  \cite{Stern2019} derived the gas cooling flow rate for a Milky Way like galaxy. 
Concisely put, their simulations are controlled numerical experiments in which gas that is initially in hydrostatic equilibrium within an external gravitational
potential is allowed to cool radiatively. They estimate the ratio between the cooling flow rate and the SFR ($\dot{M}$/SFR) to be $\sim 10$ within 100 kpc for a Milky Way like galaxy of halo mass $\sim 10^{12}$ M$_\odot$. This is broadly consistent with what we have shown are our results in Figure \ref{fig:smcooling}. Again, broadly, this estimate is consistent with results obtained utilizing
the radiatively cooling gas traced by O {\footnotesize VI} \citep{Mathews2017,McQuinn2018,Stern2018}. In particular, \cite{McQuinn2018} obtained the estimation of massive flow of cooling gas with $\sim 10-30$ M$_\odot$/yr from the O {\small VI} absorption for low redshift $L_*$ galaxies.

To pursue a more quantitative comparison and, in particular, to examine the trend in cooling rate with galaxy mass, we explore a more direct comparison to simulations. In particular, we consider the predictions of the  GAlaxy Evolution and Assembly semi-analytic model that has been developed over a series of studies \citep[GAEA;][]{delucia2007, delucia2014, hirschmann2016, xie2017, zoldan2018, fontanot2018, fontanot2020, xie2020}. It is characterised by an accurate treatment of chemical enrichment and energy recycling; a realistic stellar feedback model, a self-consistent partition of multi-phase cold gas; and an explicit treatment of the environmental effects on satellite galaxies. The model produces good agreement with the observed stellar mass function up to $z\sim 7$, and the cosmic star formation history up to $z\sim 10$ \citep{fontanot2017}. It is also able to reproduce the H{\small I} and H$_2$ fractions, as well as the quenched fractions in the local universe for central and satellite galaxies respectively \citep{xie2020}. 
In particular, the theoretical predictions in the following have been taken from the model realization applied to the Millennium Simulation \citep{springel2005} and discussed in \cite{xie2020}. 
In this semi-analytic model, baryons in galaxies are distributed in discrete components: hot gas, cold gas, stellar disk, stellar bulge, and central massive black hole. In detail,  \cite{xie2020} assume that cold gas is partitioned into H{\small I} and H$_2$, and that stars form from the H$_2$ component following the empirical law proposed by \citep{blitz2006}. This approach implies that the SFR is therefore a function of mid-plane pressure of the gas disk. The gas disk consists only of cold gas that cools either from a hydrostatic hot atmosphere when the cooling time is longer than dynamical time, or comes directly from outside of the dark matter halo when the cooling time is shorter. The former cooling regime dominates at low redshift. When cooling from a hydrostatic hot gas halo,  the cooling rate is calculated using the cooling function \citep{Sutherland1993, delucia2010}. In the latter case, all of the matter accreted into the dark matter halo cools onto the gas disk. Cooling is also allowed for satellite galaxies in GAEA.  The cold gas amassing in galactic discs is then supposed to fuel star formation. The resulting feedback reheats the cold gas at a rate that depends on the SFR, on the circular velocity and on redshift \citep{hirschmann2016}, using functional forms that have been inspired from the results of high-resolution N-body simulations \citep{Muratov2015}. Reheated gas is moved from the cold gas component to the hot gas component, leading to a complex interplay between cooling and heating. Moreover, feedback from radio-mode AGN feedback (due to inefficient gas accretion onto the central massive black holes) is assumed to provide an additional heating channel \citep{croton2006,xie2017}, that further prevents the hot gas from cooling in the more massive systems.

The resulting mass cooling rates plotted in Figure \ref{fig:smcooling} are the combination of the simulated mass of the cooling gas retained by the disk and the mass of the reheated gas. In addition, the models present calculated SFRs. The calculated SFRs are
naturally a good fit to the measured ones because the models are in part tuned to the reproduce related  constraints such as galaxy stellar masses. More impressive, however, is the independent agreement between the model predictions for the mass cooling rates and our measurements. The only significant disagreement appears at the lowest galaxy masses, where we have already noted our model may have a problem (\S\ref{sec:model}) and the GAEA model may also have a problem, partially due to limited resolution \citep{xie2017,xie2020}.

\subsection{The Feedback Mass Loading Factor}

\subsubsection{Empirical Results}

As is discussed above, the mass cooling rate is significantly higher than the SFR, demonstrating that the conversion of cooling gas into stars is inefficient and that a large fraction  of  the  gas  must  be  reheated  or  ejected. We define the mass loading factor, $\eta$, to be the ratio between the net mass cooling rate (the difference between the mass cooling rate and the SFR) and the SFR. Note that this definition may differ from that in other studies that focus on physical mass outflow. Our estimate includes the effects both of outflow and of any gas reheating that prevents that gas from participating in star formation.

The mass loading factors as calculated from our model of CGM H$\alpha$ fluxes show a rise of about a factor of 10 across galaxy mass range we explore (Figure \ref{fig:smcooling}) and in all cases have values that are significantly larger than 1. The rise is steeper at the larger M$_*$ in our sample, in the log-log plot, indicating that the dependence of $\eta$ on M$_*$ rises faster than a simple power law. The relation  in the limited mass range between $\log \eta$ and $\log \rm M_{*,10}$, where M$_{*,10}$ is again the stellar mass in units of $10^{10}$ M$_\odot$, can be approximated by a second order polynomial function as follows:

\begin{equation}
\label{eq:ml}
\log \eta = 0.66 + 0.27 \log {\rm M_{*,10} + 0.46 \log^2 M_{*,10}}
\end{equation}

\begin{figure}[htbp]
\begin{center}
\includegraphics[width = 0.48 \textwidth]{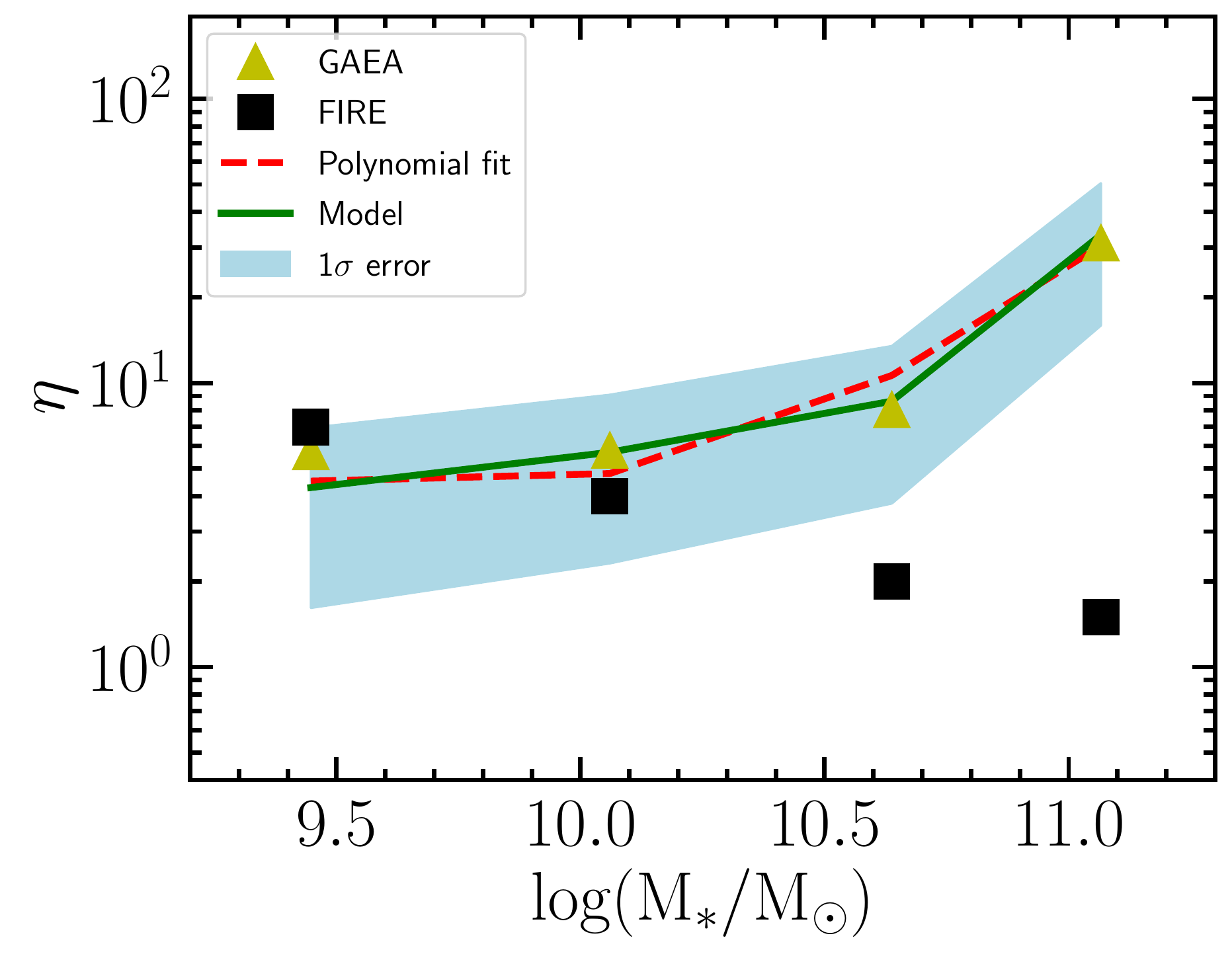}
\end{center}
\caption{The feedback mass loading factor, $\eta$, calculated using our model for $\mathcal{F}_{cool}$, with resulting systematic uncertainties plotted as the shaded region, and the results from two different published theoretical studies are presented versus primary galaxy stellar mass.  The dashed red line presents the second-order polynomial fit to  our results (Equation \ref{eq:ml}). For comparison, we  plot the mass loading factors from the GAEA  \citep{xie2020} and FIRE simulations \citep{Muratov2015,Hayward2017} as triangles and squares, respectively. The apparent discrepancy with the FIRE results is resolved in the text.}
\label{fig:massloading}
\end{figure}

\subsubsection{Comparison to Previous Theoretical  Results}

In Figure \ref{fig:massloading} we compare our estimates of the mass loading factors to two theoretical predictions. First, we compare to the same GAEA models we discussed previously. 
The agreement is excellent, which is as expected given the agreement in both the predicted vs. observed SFRs and mass cooling rates (Figure \ref{fig:smcooling}). 

Next we compare to results presented from the FIRE
(The Feedback In Realistic Environments) collaboration \citep{fire1,fire2}. They executed a series of high-resolution cosmological simulations of galaxy formation to z = 0, spanning halo
masses $\sim 10^8 - 10^{13} $ M$_\odot$, and stellar masses $\sim 10^4 - 10^{11}$ M$_\odot$.  These simulations explicitly treat the multi-phase ISM with heating and cooling physics from gas at a range of temperatures $T \sim 10 - 10^{10}$ K, 
star formation restricted only to self-gravitating, self-shielding, molecular, high density ($n_{\rm H} \gtrsim 5-50$ cm$^{-3}$) gas, and (most importantly) explicit treatment of stellar feedback including the energy, momentum, mass, and metal fluxes from SNe Types Ia \& II, stellar mass loss (O-star and AGB), radiation pressure (UV and IR), and photo-ionization and photo-electric heating. These sources of feedback reproduce
the observed relation between stellar and halo mass up to M$_{halo} \sim 10^{12}$ M$_\odot$, roughly corresponding to $M_* \sim 10^{10.5}$ M$_\odot$. Star formation rates agree well with the observed Kennicutt relation at all redshifts. 

In \cite{Muratov2015} and \cite{Hayward2017}, the investigators identified  whether a particle is considered as outflowing or inflowing by their radial velocity. Once outflowing particles are identified, they calculate outflow
rates in M$_\odot$ yr$^{-1}$ by computing the instantaneous mass flux through a thin spherical shell.  Combining the SFR from the simulation, they calculate the mass loading factors for low mass galaxies with M$_* < 10^{10}$ M$_\odot$ and upper limit for galaxies with M$_* > 10^{10}$ M$_\odot$ at redshift $\sim 0.25$, which are shown in squares in Figure \ref{fig:massloading}.  A more recent analysis with FIRE-2 simulations \citep{pandya} is able to convert these limits to measurements and those measurements lie slightly below the previous limits.
Note that these studies focus on  outflowing gas and, therefore, that the estimates of the mass loading factor, as defined here, using their results is a lower bound on what we should be measuring. 

At the low mass end with M$_* < 10^{10}$ M$_\odot$, the FIRE simulation is statistically consistent with our estimation. At M$_* > 10^{10}$ M$_\odot$, however, the FIRE simulation results significantly diverge from our estimates. However, this disagreement does not necessarily reflect a fault in either our measurements or the FIRE results.
Again recall that their 
results only represent true outflow and do not include reheating. Furthermore, their
simulations are intended to reproduce only galaxies with M$_{halo} < 10^{12}$ M$_\odot$, which for our stellar mass- halo mass relation corresponds to a stellar mass of $10^{10.5}$ M$_\odot$. Lastly, the divergence in Figure \ref{fig:massloading} is likely also reflecting the lack of feedback other than from stars in these models, e.g., AGN feedback, which is included in the GAEA modeling. The agreement among the mass loading factors from our empirical results, the semi-analytic results from GAEA, and the more detailed, physically motivated simulations from FIRE at low masses shows an encouraging convergence on our understanding of mass loading factors at those masses. 

This comparison highlights a challenge in comparing observations and theoretical results given the different definitions used for related quantities. Similarly, there are other recent studies \citep[e.g.,][]{ Mitchell2020} that present quantities that are difficult to translate into the quantities we measure. With the advent of measured H$\alpha$ fluxes, we encourage theorists to also tabulate quantities that are directly observable.

\section{Summary}
\label{sec:sum}

We present measurements of
the mass cooling rates for CGM gas and infer a feedback mass loading factors from the comparison of that rate to the measured star formation rates as a function of galactic stellar mass. 

To calculate these estimates, we first revisited our measurements of the H$\alpha$ emission line flux as a function of  stellar mass, M$_*$, and as a function of the local environment using both newer data and an improved treatment. We track the 
behavior of H$\alpha$ flux 
within two scaled annular bins around
our primary galaxies
as a function of both stellar mass ($10^{9.5} < {\rm M}_* < 10^{11}$) and scaled projected 
distance from a massive neighbor ($0.05 < R_p/R_{\rm vir} < 4$). The flux in both annular  bins drops as M$_*$ increases, indicating a strong inverse relationship between the cool gas fraction of the CGM, $f_{\rm cool}$ and M$_*$. Here $\mathcal{F}_{cool}$  describes the fraction of the cosmological baryon fraction in the halo that is in the cool ($T \sim 10^{4-5}$K) gas phase.  We also find that a strong environmental effect for the outer annular bin at the inner two smallest values of  $R_p/R_{vir}$. We fit a 
model that includes both a mass and environment dependence within the framework of the {\sl UniverseMachine} modeling \citep{Behroozi2018} and find
\begin{equation}
\mathcal{F}_{cool,{\rm isolated}}({\rm M_*}) = (0.23\pm0.01)  \left({\rm M_{*,10}}\right)^{(-0.55\pm0.03)}
\end{equation}
\noindent
for galaxies beyond the influence of a nearby, massive neighbor (as defined in \S\ref{sec:model}), and
\begin{equation}
 \mathcal{F}_{cool}(R, {\rm M_*}) =   {R\over{1.92^{+1.05}_{-0.88}
 R_{vir}}}\cdot \mathcal{F}_{cool,{\rm isolated}}
\end{equation}
\noindent
otherwise, where $R$ is the physical distance between the primary and the massive neighbor and $R_{vir}$ is the virial radius of the massive neighbor. Our prediction for the cool gas fraction is as high as 90\% for the smallest halos we probe (although we may be overestimating the fraction at the low mass end, see \S\ref{sec:model}), $23\pm 1$\% for a Milky Way like galaxy, and only a few percent for the most massive galaxies in our sample. These results, except for the lowest mass galaxies, are consistent with our previous results in Paper IV and V, despite changes to the data and treatment. The only category where there is some tension with the previous work is for the inner annular bin for the lowest mass galaxies. 

Our measurement of the H$\alpha$ flux originating from the CGM, and our models, provides us the opportunity to measure the rate at which the CGM is radiatively losing energy, and hence cooling.  We use the \cloudy modeling suite \citep{Cloudy98,Cloudy13,Cloudy17} to understand what fraction of the cooling is contributed by the H$\alpha$ flux we measure. 
In doing this, we need to account for gas at different temperatures and present arguments for why we consider only gas between  $10^4 - 10^5$ K. By enforcing a continuous cooling mass flow through the different temperatures we are able to calculate the gas mass at the different temperatures within this range and complete the calculation.

We then estimate the rate at which the CGM gas is cooling and presumably sinking into the central disk. 
We explore the dependence of the CGM emission spectra on the ionization state, the metallicity, the hydrogen density, and estimate the uncertainties in our recovered mass cooling rates.
We present both the gas mass cooling rate and effective mass loading factor required to avoid growing the disk gas mass. The mass cooling rate is a factor of several higher than the star formation rate of the central galaxies. The simple fitted relations between the mass cooling rate, MCR, and the mass loading factor, $\eta$ with galactic stellar mass are as follows: 

%\begin{equation*}
\begin{align}
\log {\rm MCR} = 0.81 + 0.48 \log {\rm M_{*,10} + 0.35 \log^2 M_{*,10}} \\
\log \eta = 0.66 + 0.27 \log {\rm M_{*,10} + 0.46 \log^2 M_{*,10}}
\end{align}
%\end{equation*}

\noindent
where M$_{*,10}$ is the galactic stellar mass in units of $10^{10}$ M$_\odot$.

Although we have used highly idealized models to transform our measured H$\alpha$ flux values into estimates of the mass cooling rates and feedback mass loading factors, it is heartening to find excellent agreement between our estimates of those quantities and those derived from two theoretical treatments that are themselves quite different in their approach \citep{xie2018,xie2020,fire1,fire2}.
The general agreement we find suggests that we have a relatively robust quantitative understanding of the cooling and feedback rates for galaxies in this mass range. The empirical confirmation of these rates provided here supports further modeling of the role of the CGM in galaxy evolution. It is perhaps somewhat surprising that our highly idealized model produces results in such good agreement given the likely complexity of CGM \citep{CGM2017}, but perhaps our measurement, which involves massively averaging over thousands and thousands of galaxy lines of sight provides a bulwark against the complexity of any individual galaxy's CGM. The field critically needs maps of the CGM in individual galaxies to further refine our understanding of this key galactic component.

\section{Acknowledgments}

DZ and HZ acknowledge financial support from NSF grant AST-1311326. KPO is funded by NASA under award No 80NSSC19K1651. XY is supported by the national science foundation of China (grant
Nos.11833005, 11890692, 11621303) and 111 project No.B20019. TF is supported by the National Key R\&D Program of China No. 2017YFA0402600.  The authors gratefully acknowledge  the SDSS III team for providing a valuable resource to the community.
Funding for SDSS-III has been provided by the Alfred P. Sloan Foundation, the Participating I institutions, the National Science Foundation, and the U.S. Department of Energy Office of Science. The SDSS-III web site is http://www.sdss3.org/.

SDSS-III is managed by the Astrophysical Research Consortium for the Participating Institutions of the SDSS-III Collaboration including the University of Arizona, the Brazilian Participation Group, Brookhaven National Laboratory, Carnegie Mellon University, University of Florida, the French Participation Group, the German Participation Group, Harvard University, the Instituto de Astrofisica de Canarias, the Michigan State/Notre Dame/JINA Participation Group, Johns Hopkins University, Lawrence Berkeley National Laboratory, Max Planck Institute for Astrophysics, Max Planck Institute for Extraterrestrial Physics, New Mexico State University, New York University, Ohio State University, Pennsylvania State University, University of Portsmouth, Princeton University, the Spanish Participation Group, University of Tokyo, University of Utah, Vanderbilt University, University of Virginia, University of Washington, and Yale University.

\bibliography{bibliography}

\end{document}